\documentstyle[aps]{revtex}
\begin{document}
\bibliographystyle{unsrt}
\title{Double Phase Slips and Spatio-Temporal Chaos\\ in a Model for 
Parametrically Excited Standing Waves}
\author{Glen D. Granzow and Hermann Riecke}
\address{Department of Engineering Sciences and Applied Mathematics\\
Northwestern University, Evanston, IL 60208, USA}
\maketitle

\begin{abstract}
We present results of numerical simulations of coupled Ginzburg-Landau 
equations that describe parametrically excited waves.  
In one dimension we focus on a new regime in which the Eckhaus sideband
instability does not lead to an overall change in the wavelength via the
occurrence of a single phase slip but instead leads to ``double phase slips''.
They are characterized by the phase slips occurring in sequential
pairs, with the second phase slip quickly following and negating the first. 
The resulting dynamics range from transient excursions from 
a fixed point resembling those seen in excitable media, to periodic solutions
of varying complexity and chaotic solutions.
In larger systems we find in addition localized spatio-temporal chaos, where 
the solution consists of a chaotic region with quiescent regions on each side.
We explain the localization using an effective phase diffusion equation which
can be viewed as arising from a homogenization of the chaotic state.
In two dimensions the double phase slips are replaced by fluctuating bound 
defect pairs.
\end{abstract}

\section{Introduction} \label{sec:intro}

Pattern formation is a striking and interesting phenomenon observed in 
many physical systems driven far from thermodynamic equilibrium.
In addition to parametrically excited waves 
(which are described below),
classic examples include the rolls formed in Rayleigh-B\'{e}nard convection
when a fluid confined between two horizontal plates is heated from below,
the annular Taylor vortices that form in a fluid between rotating 
concentric cylinders, and rotating spirals (and other patterns) 
consisting of 
regions of different chemical concentrations formed in chemical reactions such 
as the Belousov-Zhabotinsky reaction.
A review of pattern forming systems can be found in \cite{CrHo93}.
Typically, the basic equations that describe these pattern-forming 
experiments are quite complicated (such as the Navier Stokes equations).
However, for pattern-forming systems near onset, the complicated system of 
equations can often be simplified using asymptotic expansions for small 
amplitudes. 
The resulting amplitude equations (e.g. \cite{Ma90,GoSt88,NeWh69,Se69}) describe 
the variation of the pattern on large space and time scales. 

In many systems transitions from simple to chaotic behavior 
are found when a control parameter 
(such as the temperature difference in Rayleigh-B\'{e}nard convection, 
the angular velocity of the inner cylinder in Taylor-Couette flow, or the 
supplied concentration of a particular chemical in a reaction diffusion system)
is changed.
Studies of chaos began 
with low-dimensional attractors as found, for instance, in the Lorentz system 
\cite{Lo63} and have
progressed to chaotic behavior in both space and time in spatially extended
systems with a large number of degrees of freedom.
Of particular interest in both experimental and mathematical systems is 
identifying features that allow the different types of observed chaos to be 
classified in some way.  
A well studied example is 
the complex Ginzburg-Landau equation, for which a number of different chaotic
regimes have been identified \cite{ShPu92,ArKr94,EgGr95,ChMa96,MaCh96}.  
Relevant for the present paper is the observation that a distinction can be
drawn between a phase-chaotic regime and amplitude-chaotic ones.
In one dimension amplitude chaos is characterized by the occurrence of phase 
slips during which the amplitude of the wave goes to zero and the total phase of 
the system changes by $2 \pi$, that is, a wavelength is inserted or eliminated.
In phase chaos there are essentially no phase slips \cite{EgGr95,MaCh96}
and the system can be described by an equation for the phase of the wave alone 
\cite{Ku78}.
For the complex Ginzburg-Landau equation this phase equation is the 
Kuramoto-Sivashinsky equation.
Phase equations break down during phase slips so they cannot 
be used to describe amplitude chaos.

In this paper we consider parametrically excited waves, 
which are also a classic pattern forming system \cite{Fa31}. 
Parametrically excited waves arise quite generally in systems exhibiting 
weakly damped modes that are oscillatory in both space and time, when 
these modes are forced at twice their natural frequency.
Faraday's experiment \cite{Fa31} in which surface waves are formed on a 
liquid in a
vertically oscillated container is probably the most well known example.
Stripe, square, and hexagon patterns (e.g. \cite{KuGo96a}), as well as 
quasipatterns \cite{EdFa94}, spirals \cite{KiKo96}, and chaotic states 
(e.g. \cite{TuRa89,KuGo96,KuGo96a}) 
have all been observed in Faraday experiments with various fluids, forcing
amplitudes, and frequencies.
Parametrically excited waves have also been observed in magnetic materials 
(spin waves) \cite{Su57,El93},
in electroconvection in nematic liquid crystals \cite{ReRa88,ToRe90}, 
and in Taylor-Dean flow \cite{TeAn96}.

In this paper we discuss dynamics arising in numerical simulations of a system 
of equations that model parametrically excited waves and focus on solutions 
which are chaotic in space and time.
In our simulations we find chaotic solutions which are characterized 
by a new phenomenon: double phase slips in one dimension and the
analogous bound defect pairs in two dimensions.
These phenomena are described in sections \ref{sec:phaseslips} and 
\ref{sec:twoD}.
In addition to chaotic solutions 
where these phenomena occur irregularly in space and time, 
which are described in sections \ref{sec:complex1D} and \ref{sec:extended1D}, 
we also describe transient and periodic solutions involving double phase slips
in sections \ref{sec:transient1D} and \ref{sec:periodic1D}. 

Of central interest in our one-dimensional simulations are localized
chaotic states (section \ref{sec:localized1D}).
That is, there are solutions that include a spatial region, bounded by 
quiescent regions on each side, in which the dynamics are irregular in 
space and time.
Experimentally, similar phenomena have been observed in Taylor vortex
flow  \cite{BaAn86}, Rayleigh-B\'{e}nard convection \cite{CiRu87}, 
and parametrically excited surface waves \cite{KuGo96,KuGo96a}.
The localization mechanism in these experimental systems is only poorly
understood.
For the system of equations discussed in this paper the localization
can be understood using an effective phase diffusion equation.
We confirm the validity of this approach by explicitly determining the
effective phase diffusion coefficient in section \ref{sec:effectiveD}.
A short account of these results has been presented previously \cite{GrRi96}.

While most of the results in this paper are for a one dimensional system,
section \ref{sec:twoD} describes results for an analogous system in two 
dimensions.
Section \ref{sec:conclusions} concludes our paper with a summary and discussion 
of the results.

\section{The Model} \label{sec:model}

The equations we consider in this paper generically describe a system with
a uniform state that has a supercritical Hopf bifurcation to a state that
is periodic in both space and time when subjected to a uniform, 
constant forcing. 
Subjecting the system to an additional spatially uniform, time-periodic forcing
at approximately twice the Hopf frequency
creates a codimension-three bifurcation point involving the following
three parameters:  
1) the amplitude of the time-independent forcing,  
2) the amplitude of the periodic forcing, and 
3) the frequency of the periodic forcing.
In the vicinity of this codimension-three bifurcation point the following 
amplitude equations \cite{RiCr88,Wa88,RiCr91} are derived by considering the 
reflection and translation symmetries in space and the weakly broken 
translation symmetry in time:
\begin{eqnarray}
\partial_TA+s\partial_XA &=& d\partial_X^2A+aA+bB
 +cA(|A|^2+|B|^2)+gA|B|^2,                               \label{oneDA} \\
\partial_TB-s\partial_XB &=& d^*\partial_X^2B+a^*B+bA
 +c^*B(|A|^2+|B|^2)+g^*B|A|^2.                           \label{oneDB}
\end{eqnarray}
The dependent variables $A$ and $B$ are complex, and represent 
amplitudes of left and 
right traveling waves that are summed together to yield the solution of the 
underlying system:
\begin{equation}
u(x,t)=\epsilon A (X,T)e^{i(q_cx-\frac{\omega_e}{2}t)} 
      +\epsilon B (X,T)e^{i(q_cx+\frac{\omega_e}{2}t)}
      +c.c.+o(\epsilon),                                 \label{oneDu}
\end{equation}
where $u(x,t)$ would give, for example, the position of the fluid surface in the
Faraday experiment.
The amplitudes $A$ and $B$ vary on the slow time and space scales, 
$T=\epsilon^2t$ and $X=\epsilon x$.
The coefficients in (\ref{oneDA},\ref{oneDB}) are 
complex except for $s$ and $b$, which are real.
The real part of the coefficient of the linear term $a$ gives the linear 
damping of the traveling waves in the absence of the periodic forcing and is
proportional to the distance from the Hopf bifurcation. 
The coefficient of the linear coupling term $b$ gives the amplitude of the 
periodic forcing.
(Note that this term breaks the time translation symmetry.)
The imaginary part of the coefficient of the linear term $a$ gives the 
difference between the frequency of the unforced waves and half the forcing 
frequency $\omega_e$.

While in general a codimension-three bifurcation might be difficult to realize
in an experiment, in the above situation 
(involving a system that undergoes a Hopf bifurcation) 
each of the three bifurcation parameters are natural control parameters in the
system.
In the Faraday experiment, however, there is no Hopf bifurcation and the
above equations are only valid in the limit of weak damping (small viscosity).
In this case the coefficients of the nonlinear terms $c$ and $g$ would be
purely imaginary, since their real parts also represent (nonlinear) damping
and would be higher order.
In addition, for the Faraday experiment the group velocity parameter $s$
would generally be complex, and $s$ in (\ref{oneDB})
would be replaced by its complex conjugate $s^*$.

There is one caveat concerning (\ref{oneDA},\ref{oneDB}) when they
are used to describe either the Faraday experiment or parametrically excited
waves induced near a Hopf bifurcation point.
With the scaling $X=\epsilon x$, the group velocity term 
(involving $\partial_X$) would generically appear at lower order, so 
(\ref{oneDA},\ref{oneDB}) require that the group velocity is small.
If it is not small a strict asymptotic analysis would require the introduction 
of a second slow time scale and would lead to equations in which the 
cross-coupling between $A$ and $B$ arises through a spatially averaged term
\cite{KnDe90,Kn92,PiKn94a,MaVo92,Ve93}.

Note that (\ref{oneDA},\ref{oneDB}) are in one (spatial) dimension.  
With periodic
boundary conditions (as will be imposed in our numerical simulations)
these equations 
would describe (for example) the Faraday experiment in an annular container. 
Equations in two (spatial) dimensions (see (\ref{twoD1},\ref{twoD2}) below) are 
discussed in section \ref{sec:twoD}.  

Solutions of (\ref{oneDA},\ref{oneDB}) of the form $A=A_0e^{iQX}$,
$B=B_0e^{iQX}$ include those where $A_0$ and $B_0$ are constants of unequal 
magnitude that correspond to traveling waves in the underlying system 
(see (\ref{oneDu}))
, and those where
$A_0$ and $B_0$ are constants with equal magnitude that correspond to standing 
waves in the underlying system \cite{RiCr88,Wa88}.
It is these standing-wave solutions, which are phase locked to the periodic 
forcing, that are the focus of this paper. 
In particular, we discuss the ensuing dynamics when these standing waves are 
perturbed.
There are also solutions where $A_0$ and $B_0$ have equal magnitude 
but vary periodically in time, which correspond to standing waves that are not 
phase locked to the external forcing \cite{RiCr88}.  
We do not treat these standing waves in this paper.

The standing waves which are phase locked to the external forcing are 
stationary solutions of (\ref{oneDA},\ref{oneDB}) and are given by 
\cite{RiCr88,Wa88}
\begin{eqnarray}
A & = & Re^{i(QX+\phi_A)}           \label{eqn:solnA} \\
B & = & Re^{i(QX+\phi_B)}           \label{eqn:solnB} \\
\mbox{where: \hspace{2ex} } 
R^2 & = & \frac{-(\hat{a}_r n_r + \hat{a}_i n_i) \pm \sqrt{(\hat{a}_r n_r 
+ \hat{a}_i n_i)^2-(n_r^2+n_i^2) (\hat{a}_r^2+\hat{a}_i^2-b^2)}}
{(n_r^2+n_i^2)}                     \label{eqn:R} \\
\phi_B-\phi_A & = & \arctan \frac{\hat{a}_i+n_iR^2}{\hat{a}_r+n_rR^2} \\
\mbox{and: \hspace{2ex} } 
\hat{a} & \equiv & \hat{a}_r+i\hat{a}_i \hspace{1ex} = \hspace{1ex} a-isQ-dQ^2 \\
n & \equiv & n_r + in_i \hspace{1ex} = \hspace{1ex} 2c+g.  \label{eqn:definen}
\end{eqnarray}
They arise from the trivial solution in a steady bifurcation at the onset
$b^2=\hat{a}^2_r+\hat{a}^2_i$.
Equations (\ref{eqn:solnA},\ref{eqn:solnB}) actually give a family of solutions 
for each wavenumber $Q$ since the phase $\phi_A$ (or $\phi_B$)
can be chosen arbitrarily.  
Changing the value of $\phi_A$ (with $\phi_B-\phi_A$ fixed) corresponds to 
spatial translations of the solution.

Close to onset  
(\ref{oneDA},\ref{oneDB}) can be reduced to a single Ginzburg-Landau
equation with real coefficients \cite{Ri90a}.  
This implies that the waves are stable near onset in a band of wavenumbers
which is limited by the Eckhaus (i.e. a long-wavelength) instability.
Near onset the neutral and Eckhaus curves
are parabolic \cite{Ec65}.  
For larger values of the forcing amplitudes $b$, the Eckhaus curve need 
no longer be parabolic and can assume a variety of shapes.
For $a=-0.05$, $c=-1+4i$, $d=1+0.5i$, $s=0.2$, $g=-1-12i$, the Eckhaus curve 
closes on itself as shown in Figure \ref{fig:linstab} \cite{Ri90a}.
Thus, for $b>0.85$ all standing waves become unstable, somewhat reminiscent
of the Benjamin-Feir instability of traveling waves \cite{BeFe67}.
In this paper we focus on the dynamics resulting from the Eckhaus instability.
All computations presented in this paper were made using the same parameter 
values as used in Figure \ref{fig:linstab}.

In reference \cite{Ri90a} the Eckhaus curve in Figure \ref{fig:linstab}
was obtained using phase dynamics.
A more complete linear stability 
analysis considering sideband perturbations with arbitrary wavenumbers confirms 
that there is no short-wavelength instability for these parameter values.

\section{Phase Slips and Double Phase Slips} \label{sec:phaseslips}

The main questions to be addressed in this paper are:  
What are the solutions to
(\ref{oneDA},\ref{oneDB}) if started with initial conditions that 
correspond to unstable standing waves?
What features characterize the solutions?
How do the solutions differ as one changes the forcing amplitude $b$ and
the wavenumber of the initial condition $Q$?
We investigate these questions by calculating numerical approximations to the 
solutions
of (\ref{oneDA},\ref{oneDB}) using a pseudospectral method in space
(periodic boundary conditions) and a fourth-order Runge-Kutta/integrating-factor
scheme in time.

The dynamics described in this paper all incorporate
a feature we call a ``double phase slip''.
In this section we describe this feature and contrast it to an ordinary single 
phase slip, which is a well known feature of 
many systems (including the solution of a single Ginzburg-Landau
equation with real coefficients) when started with an initial condition that
is Eckhaus unstable.
In order to excite the Eckhaus instability we typically choose an initial
condition that is a wave, perturbed so that the local wavenumber
varies slightly throughout the domain such that it has a maximum and a minimum 
at distinct locations.

Since near onset (\ref{oneDA},\ref{oneDB}) can be reduced to a single
real Ginzburg-Landau equation \cite{Ri90a}, the solution of 
(\ref{oneDA},\ref{oneDB}) for small values of $b$ should resemble
the solution of a single real Ginzburg-Landau equation.
This is indeed seen in our simulations.
For small values of $b$ ($b=0.1$, say) and an initial condition slightly to the
right of the Eckhaus stable region in Figure \ref{fig:linstab}, the solution 
undergoes a phase slip near the location of the initial maximum in the 
wavenumber 
that reduces the wave number and moves the solution into the stable region.
This behavior is shown in Figure \ref{fig:phaseslip}a.
For larger values of $b$ however ($b=0.6$, say), the behavior of the solution is
markedly different, as shown in Figure \ref{fig:phaseslip}b; 
given an initial condition slightly to the right of the
Eckhaus stable region, the solution undergoes a phase slip that reduces the
average wave number but a short time later undergoes a second phase slip at
essentially the same location in space that restores the wave number to its
original value.  This {\it double phase slip} causes the 
solution to remain in the Eckhaus unstable region, thus allowing persistent
dynamics.

In our simulations a phase slip in $A$ is always accompanied by an essentially 
coincident phase slip in $B$.
Thus when we refer to a double phase slip in the solution of 
(\ref{oneDA},\ref{oneDB}) we mean that there are double phase slips in both 
$A$ and $B$ at approximately the same point in space and time.

Double phase slips occur in a variety of solutions to 
(\ref{oneDA},\ref{oneDB})
including periodic solutions, chaotic solutions, and non-trivial transients.
These solutions are described in the following sections.
We first present results for small system sizes.
Then we discuss large systems and show that they allow a simplified long-wave
description.

\section{Dynamics of a Small System} \label{small1D}

For the simulations presented in this section the domain size $L$ has been 
chosen such that the initial conditions, which have an average wave number in 
the vicinity of the Eckhaus curve, include only five wavelengths.
Three different types of solutions are presented:  
1) Transient solutions where there is only a single double phase slip, 
2) Solutions where double phase slips occur periodically in time, and 
3) Complex solutions where double phase slips occur irregularly in space and 
time.
Figure \ref{fig:solutions} shows where examples of these solutions occur in 
relation to the Eckhaus curve that is shown in Figure \ref{fig:linstab}.
The initial conditions for these solutions are close to the standing wave 
solutions but perturbed such that there is a maximum in the local wave number
that triggers the first double phase slip and determines its location.
Examples of each of the solution types are described below.

\subsection{Transient Solutions} \label{sec:transient1D}

In the transient solutions described here,
there is only one double phase slip.
After the double phase slip, the solution relaxes to the stationary solution
corresponding to the regular standing wave state in the underlying system.
These transient solutions resemble the response of an excitable medium;
beginning with an initial condition that is in the vicinity of  
a stable fixed point, the transient solution makes an excursion far away from
the fixed point before finally coming to rest there.
This is most clearly seen in the phase-plane representation of the solution
shown in Figure \ref{fig:phaseplane1DPS}. 
It shows the magnitudes of the
(spatial) Fourier modes corresponding to 4 and 5 wavelengths in the domain.
Each dot in Figure \ref{fig:phaseplane1DPS} represents an instant in time.
The square indicates the initial condition which is the perturbed 5-wave state.
The loop in Figure \ref{fig:phaseplane1DPS} is traversed once, with the final 
state of the solution being the fixed point on the vertical axis (marked by a 
diamond) corresponding to the (stable) 5-wave solution.

Solutions with a single double phase slip occur only where the 
initial condition is a perturbation of a {\it stable} solution.
The lowest point on Figure \ref{fig:solutions} appears to contradict this 
statement since it appears on the outside of the Eckhaus curve.
However, this is due to the fact that the Eckhaus curve represents a 
long-wavelength instability.
Since arbitrarily small wavenumbers are not present in our finite domain the 
true stability curve for our finite system is slightly outside the Eckhaus 
curve and all of the solutions with a single double phase slip are in fact
within the stable region.
At the stability limit an unstable saddle-point solution branches 
subcritically off the branch of 
solutions given by (\ref{eqn:solnA}-\ref{eqn:definen}) \cite{KrZi85,TuBa90}.
The perturbation required to excite a double phase slip must place the initial
condition beyond this saddle point in phase space.
As the wavenumber is moved further into the Eckhaus stable region,
the bifurcated saddle point moves further from the stable stationary solution
in phase space, 
and a larger perturbation is required to excite the system.

\subsection{Periodic Solutions} \label{sec:periodic1D}

In the periodic solutions, double phase slips occur at the same locations
in space, periodically in time.
An example of the simplest of these periodic solutions is shown in 
Figure \ref{fig:simpleperiodic} where the location in space and time of each 
double phase slip is indicated by a circle.
The time between each double phase slip is much larger than the time between
the individual phase slips which make up a double phase slip.
This can be appreciated by looking at Figure \ref{fig:qvstime} which shows the 
(spatial) average of the wavenumber plotted versus time.
Each of the downward ``spikes'' in Figure \ref{fig:qvstime} represents
a double phase slip; the average wavenumber is reduced by the first phase
slip, then restored by the second.
Figure \ref{fig:phaseplaneSPS} shows the phase-plane representation of the 
simple periodic solution. 
The loop in Figure \ref{fig:phaseplaneSPS} is traversed periodically, with the 
upper left portion of the loop 
(where the mode 5 magnitude is large and the mode 4 magnitude is small)
being passed between double phase slips, and the lower right portion of the loop 
(where the mode 5 magnitude is smaller and the mode 4 magnitude is larger)
being passed during a double phase slip.

Although the periodic solutions occur for larger wavenumbers (for a given value 
of $b$) than the transient solutions where there is only one double phase slip 
(see Figure \ref{fig:solutions}) we find that the periodic solutions can still 
occur within the Eckhaus stable region
indicating that stable stationary solutions and stable periodic solutions can 
coexist.

As the wavenumber is decreased the period of the simple periodic solutions
diverges as shown in Figure \ref{fig:period}.
This appears to be due to the 
unstable saddle-point solution which branches subcritically off the branch of 
solutions 
given by (\ref{eqn:solnA}-\ref{eqn:definen})
at the Eckhaus stability limit.
Inside the Eckhaus stable region,
as the average wavenumber of the initial condition is decreased by increasing
the size of the system $L$, the saddle
point moves away from the stable fixed point and towards the periodic
orbit in phase space.
This causes the periodic orbit to pass closer to the saddle point, increasing
its period (cf. Figure \ref{fig:period}).
Our simulations suggest that
the divergence of the period occurs when the periodic orbit becomes a
homoclinic orbit of the saddle point.
Beyond this point the periodic solutions no longer exist and the excitable
transient solutions occur.

Increasing the wavenumber from those which produce simple periodic 
solutions we find a period doubling bifurcation which leads to the 
solution shown in Figure \ref{fig:perioddoubled}.
Increasing the wavenumber further results in even more complicated 
periodic (or quasiperiodic) states such as that shown in Figure 
\ref{fig:travelingwave} 
where a wave of double phase slips propagate through the system.
The exact form of the more complicated periodic solutions may depend on the
value of $b$ but the trend of simple periodic solutions losing stability to
more complicated periodic solutions was observed for all investigated values 
of $b$ that exhibit periodic solutions.

\subsection{Complex Solutions} \label{sec:complex1D}

Figure \ref{fig:complex} shows a complex solution where the double phase slips 
occur irregularly in space and time.
The complex solutions occur for even larger wavenumbers 
(for a given value of $b$) than the periodic solutions.
If the average wave number of the initial condition is too large however, 
a single phase slip occurs, 
which reduces the wavenumber.
The phase slip need not take the solution all the way into the stable band.
Instead the solution can remain in the domain of periodic or even complex 
dynamics.


\section{Dynamics of a Large System} \label{sec:large1D}

In this section we present solutions for larger domains than the solutions 
discussed in the previous section.
For the particular solutions discussed, approximately 50 wavelengths with a 
wavenumber near the 
Eckhaus curve fit in the domain whereas in the previous section only 5 did.
We focus on the complex solutions where the double phase slips occur irregularly
in space and time.
Based on the irregular appearance of these solutions we believe them to be chaotic.
Autocorrelation functions calculated using the location of the double phase 
slips show
that these locations are not correlated for large distances and times, 
confirming the chaotic nature of the solutions in both space and time.
We discuss two different solution types:
1) Extended spatio-temporal chaos where the double phase 
slips occur irregularly across the entire domain, and 
2) Localized spatio-temporal chaos where the irregular occurrence of double 
phase slips is restricted to only part of the spatial domain.

\subsection{Extended Spatio-Temporal Chaos} \label{sec:extended1D}

Figure \ref{fig:extended}a shows a space-time diagram similar to Figure 
\ref{fig:complex} for an extended chaotic solution in a large system,
with each double phase slip represented by a single dot. 
The extended chaotic states exist over a range of wavenumbers with the
space-time density of double phase slips increasing with wavenumber as seen in 
Figure \ref{fig:extended}b.
Beyond a certain wavenumber the chaotic state loses stability, however.
A solution with such a large initial wavenumber is represented 
in Figure \ref{fig:extended}c where at
$t\approx13000$ and $x\approx265$ a {\it single} phase slip occurs, which
reduces the local wavenumber.
Subsequently there are no phase slips in this area until the higher
wavenumber in the regions on either side diffuses in.
Eventually the wavenumber and the density of double phase slips in Figure 
\ref{fig:extended}c become homogeneous again and their final values are the 
same as those in Figure \ref{fig:extended}b.
The transition appears to be related to the fact that
with the large wavenumber (that is, before the single phase slip in Figure
\ref{fig:extended}c) the time between the occurrence of
double phase slips at a given location in space is similar to the duration of
each double phase slip, i.e. the time between 
the two phase slips within a double phase slip.
This is due to both the high density of double phase slips and their tendency to 
reoccur at the same location in space.

\subsection{Localized Spatio-Temporal Chaos} \label{sec:localized1D}

Perhaps the most interesting solutions arising in our simulations are those
featuring localized spatio-temporal chaos.
An example is shown in Figure \ref{fig:localized}.
As in Figure \ref{fig:extended}, each double phase slip is represented by a 
single dot.
We start with an initial condition with a maximum in the local
wavenumber that triggers an initial double phase slip.
This initial double phase slip triggers additional double phase slips in the
vicinity of the first.
Originally the size of the region in which the double phase slips occur grows
with time.  
However, for later time, the size of the region ceases to grow and the chaos is
seen to be localized.
Outside of the region in which the double phase slips occur, the solution is the
time-independent solution which corresponds to standing waves in the underlying
system.

Figure \ref{fig:width} shows the width of the chaotic region as a function of 
time for different initial widths bracketing the final width of the state shown 
in Figure \ref{fig:localized}.
For both initial conditions the width converges to the same final value
indicating that in the final state the spatio-temporal chaos is indeed 
localized.

\subsection{Effective Phase Equation} \label{sec:effectiveD}

The localization mechanism for the steady and chaotic regions in 
Figure \ref{fig:localized} can be understood using an effective phase 
diffusion equation.
In this section we present this analysis.
Our analysis uses the phase of $A(X,T)$ but an analysis based on $B(X,T)$
would yield the same results.

From our investigation of solutions with extended chaos we expect the
chaotic region to have a larger local wavenumber than the region where the
solution is time independent 
(see Figure \ref{fig:solutions}).
Figure \ref{fig:wavenumber} shows the local wavenumber of the time average of 
the solution $A(X,T)$ 
(averaged from $t=2,000,000$ to $t=2,010,000$) for the state represented in 
Figure \ref{fig:localized} and confirms this;
the local wavenumber in the chaotic region is larger than the local
wave number in the quiescent region.
The local wavenumber for $B(X,T)$ is similar.
So the steady and chaotic regions in Figure \ref{fig:localized} can be
viewed as domains with differing wavenumber.

Somewhat similar stationary states consisting of domains of differing wavenumber 
have been observed experimentally in convection in narrow channels 
\cite{HeVi92}, as well as in Taylor vortex flow \cite{BaAn86} and in 
two-dimensional optical patterns \cite{ReRa96}.
The convection states can be described by the phase diffusion equation
\footnote{Certain phenomena such as the locking of adjacent domain walls into
each other are not captured by the phase equation (\ref{eqn:diffusion}).
For small amplitudes of the pattern the locking can be described by suitable
Ginzburg-Landau equations \cite{RaRi95a,RaRi96}.}
\cite{BrDe89,Ri90a,RaRi95a,RaRi96}
\begin{equation} 
\partial_T \phi (X,T) = D(q) \partial_{X}^{2} \phi (X,T) + h.o.t.,  
                                                           \label{eqn:diffusion}
\end{equation}
which is nonlinear due to the dependence of $D$ on 
$q \equiv \partial_X \phi$.
The higher order terms include higher 
derivative terms (e.g. $G\partial_X^4 \phi(X,T)$) which are important when
D is small or negative, or when these higher derivatives are large.
Stable solutions of (\ref{eqn:diffusion}) 
that consist of domains of differing wavenumber can occur
if the diffusion coefficient $D(q)$ is negative only over a small range of 
wavenumbers.
In this situation, initial conditions with a uniform wavenumber in this unstable
range will evolve to a structure consisting of domains in which the local
wavenumber lies in either of the two adjacent ranges of stable wavenumbers.
At the interface between these domains, the higher-order terms in  
(\ref{eqn:diffusion}) are
important and there is an internal transition layer 
(similar to a boundary layer).
These transition layers can be studied using matched asymptotic expansions as 
has been done for the Cahn-Hilliard equation and other related equations 
similar to (\ref{eqn:diffusion}) \cite{Wi96}.

Equation (\ref{eqn:diffusion}) is only valid as long as the gradients in the 
wavenumber are small.
During phase slips this is not the case.
In fact, the phase $\phi$ is not even defined at one instant during the phase
slip process.
The solution shown in Figure \ref{fig:localized} includes phase slips and thus 
cannot be described by (\ref{eqn:diffusion}).
However, because each phase slip in the solution in Figure \ref{fig:localized} 
is immediately followed by a second phase slip
which restores the phase to its value before the first phase slip, on a long 
time scale the effective total phase $\Delta\hat{\phi}=\int_0^L \hat{q} dX$ 
of the system is conserved.
Here the effective wavenumber $\hat{q}$ is defined to be 
$\partial_X \hat{\phi}$ where $\hat{\phi}$ is the phase of a running time 
average of $A(X,T)$:
\begin{equation}
\hat{A}(X,T)=\frac{1}{\Delta T}\int_{T-\Delta T}^{\Delta T} A(X,T) dT
\equiv \hat{R}e^{i\hat{\phi}}.  \label{eqn:Ahat}
\end{equation}
The time interval $\Delta T$ is chosen so that (\ref{eqn:Ahat}) averages over
many double phase slips.
An equivalent analysis could also be based on the phase of a running time average 
of $B(X,T)$ instead of $A(X,T)$.
The $A(X,T)$ portion of the solution to (\ref{oneDA},\ref{oneDB}) can be 
written as
\begin{equation}
A(X,T)=\hat{A}(X,T)+F_A(X,T)
\end{equation}
where $F_A(X,T) \equiv A(X,T)-\hat{A}(X,T)$  
contains the rapid fluctuations due to the phase slips and
has small magnitude except in the neighborhood of a phase slip. 
$\hat{A}(X,T)$ contains variations on slower time and space scales and we 
assume that $\hat{R}=\hat{R}(\hat{q})$ (this is similar to (\ref{eqn:R})).
Due to the translation symmetry in time and 
translation and reflection symmetries in space of (\ref{oneDA},\ref{oneDB}),
the long-wave evolution equation for the effective phase $\hat{\phi}(X,T)$ is 
expected to be a diffusion equation similar to (\ref{eqn:diffusion})
\begin{equation} 
\partial_T \hat{\phi} (X,T) = \hat{D}(\hat{q}) \partial_{X}^{2} \hat{\phi} (X,T) 
+ h.o.t..     \label{eqn:effectiveD}
\end{equation}
In order to test this explicitly, we measure the effective phase
diffusion coefficient $\hat{D}$ by measuring the response of the
effective phase of the chaotic state to a localized time-periodic forcing.
This approach to measuring the effective diffusion coefficient of a chaotic 
state is similar to that used in experiments on turbulent Taylor vortex
flow wherein an endcap was moved sinusoidally \cite{WuAn92}.
Perhaps it should be emphasised that, while we want to measure the effective
diffusion coefficient to understand the localized chaotic states (as shown in 
Figure \ref{fig:localized}), the actual measurements described in this section 
are done on extended chaotic states (as shown in Figure \ref{fig:extended}).
Localized forcing of the effective phase in our simulations is accomplished by 
including a spatially local, time-periodic advection term in 
(\ref{oneDA},\ref{oneDB}) giving:
\begin{eqnarray}
\partial_TA+(v+s)\partial_XA &=& d\partial_X^2A+aA+bB +cA(|A|^2+|B|^2)+gA|B|^2
       \label{eqn:advectA} \\
\partial_TB+(v-s)\partial_XB &=& d^*\partial_X^2B+a^*B+bA +c^*B(|A|^2+|B|^2)+g^*B|A|^2
       \label{eqn:advectB} \\
\mbox{with $v$} &=& 
\left\{ 
\begin{array}{ll} 
v_{0}\,\sin(\omega_0 T) & \mbox{for $X_1 \leq X \leq X_2$} \nonumber \\
   0                  & \mbox{otherwise.} \nonumber 
\end{array}
\right.
\end{eqnarray}
For small values of $v_0$ and $\omega_0$, the localized advection term causes formerly 
stationary solutions to drift 
within the region $X_{1} \leq X \leq X_{2}$, periodically reversing their direction
according to the sign of $v$.
For the regions $X < X_{1}$, and $X > X_{2}$ the situation resembles an 
imposed boundary
condition at $X = X_{1}$, and $X = X_{2}$ at which the phase of the solution
varies sinusoidally.

The solution to (\ref{eqn:effectiveD})
for $X>X_2$, which satisfies the boundary condition
$\hat{\phi} (X_2,T) = \epsilon\, \sin(\omega_0 T)$, is 
\begin{eqnarray}
\hat{\phi} (X,T) &=& \hat{Q}_0(X-X_2)+\epsilon e^{- \alpha (X-X_2) } 
\sin(\omega_0 T - \beta (X-X_2))                    \label{eqn:solutionphihat} \\
\mbox{where $\alpha$} &=& \beta = \sqrt{\frac{\omega_0}{2\hat{D}}}. \label{eqn:alphabeta}
\end{eqnarray}
If the effective phase of the solution $A(X,T)$ in the region 
$X > X_{2}$ obeys this relationship we have 
\begin{eqnarray}
\hat{A}(X,T) & = & \hat{R} e^{i\hat{\phi}(X,T)} 
               =   \hat{R} e^{i(\hat{Q}_0(X-X_2)+\tilde{\phi})} \\
\mbox{where \hspace{2ex}} \tilde{\phi} 
       & = & \epsilon e^{- \alpha (X-X_2) } \sin(\omega_0 T - \beta (X-X_2)) \\
\mbox{and \hspace{2ex}} \hat{R} 
       & = & \hat{R}(\partial_X \hat{\phi}(X,T)) = \hat{R}(\hat{Q}_0+\partial_X\tilde{\phi}).
\end{eqnarray}
If $v_0$ and $\omega_0$ are small then the magnitude of the periodic variations 
in the effective phase $\tilde{\phi}$ will also be small.
Taylor expanding $\hat{R}(\hat{Q}_0+\partial_X\tilde{\phi})$ 
and $e^{i\tilde{\phi}}$ gives
\begin{equation}
\hat{A}(X,T) \sim (\hat{R}(\hat{Q}_0)+\hat{R}^\prime(\hat{Q}_0)\partial_X \tilde{\phi}+\ldots)
e^{i\hat{Q}_0(X-X_2)}(1+i\tilde{\phi}+\ldots).
\end{equation}
Taking the Fourier transform of $A(X,T)$ in time,
\begin{equation}
{\cal A}(X,\Omega)=\frac{1}{\tau}\int_0^\tau e^{i\Omega T}A(X,T) dT
 =\frac{1}{\tau}\int_0^\tau e^{i\Omega T}\hat{A}(X,T) dT
 +\frac{1}{\tau}\int_0^\tau e^{i\Omega T}F_A(X,T) dT,
                                           \label{eqn:fourierintegral}
\end{equation}
and assuming that the fast variations $F_A(X,T)$ give only negligible 
contributions to the Fourier components at low frequencies yields
\begin{eqnarray}
{\cal A}(X,-\omega_0) & \approx & \frac{\epsilon}{2}
\left[( \hat{R}(\hat{Q}_0)-\beta \hat{R}^\prime(\hat{Q}_0))+i \alpha \hat{R}^\prime(\hat{Q}_0) \right]  
e^{-\alpha (X-X_2)+i(\hat{Q}_0-\beta)(X-X_2)}  \label{eqn:fourierminus}\\
{\cal A}(X, \omega_0) & \approx & \frac{\epsilon}{2}
\left[(-\hat{R}(\hat{Q}_0)-\beta \hat{R}^\prime(\hat{Q}_0))-i \alpha \hat{R}^\prime(\hat{Q}_0) \right]  
e^{-\alpha (X-X_2)+i(\hat{Q}_0+\beta)(X-X_2)}. \label{eqn:fourierplus}
\end{eqnarray}
Thus, the magnitude of the
Fourier modes ${\cal A}(X,-\omega_0)$ and ${\cal A}(X,\omega_0)$
(corresponding to the frequency $\omega_0$ 
in the localized advection term)
will decay exponentially in space as $X-X_2$ increases 
if the phase behaves diffusively.
Figure \ref{fig:alpha} shows that this exponential behavior is realized in 
our simulations.
It gives the absolute value of the Fourier mode 
${\cal A}(X,\omega_0)$
versus space for two different values of the frequency $\omega_0$.
According to (\ref{eqn:alphabeta}) the decay rate $\alpha$ is 
proportional to $\omega_0^{1/2}$ for diffusive behavior.
The ratio of the slopes of the curves in Figure \ref{fig:alpha} are consistent 
with this diffusive scaling.
Note that for larger frequencies, and therefore larger gradients of 
$\hat{\phi}$, higher order terms  
(like $-\hat{G} \partial_X^4 {\hat \phi}$, etc.) 
have to be kept in the diffusion equation which affect the dependence on 
$\omega$.
For $\hat{G}>0$, $\alpha$ is expected to go to zero slower than $\omega^{1/2}$:
\begin{equation}
\frac{\alpha(\omega_1)}{\alpha(\omega_2)} = 
\left(\frac{\omega_1}{\omega_2}\right)^{1/2}[1+\frac{\hat{G}}{2\hat{D}^2}
(\omega_2-\omega_1)+ O(\omega_1\omega_2)].
\end{equation}
We found analogous behavior in the nonchaotic regime where our simulations
were able to reproduce the analytical value of the diffusion coefficient.

The phase slips introduce noise which can be seen in Figure \ref{fig:alpha}. 
In order to reduce the effect of this noise and extract a reliable decay rate 
$\alpha$, the Fourier integral (\ref{eqn:fourierintegral}) has been 
extended over many periods 
($\tau=88 \times 2\pi/\omega_0$) 
of the function $v(X,T)$ in the advection term 
in (\ref{eqn:advectA},\ref{eqn:advectB}).

From the decay rate $\alpha$ 
we can calculate the diffusion coefficient $\hat{D}$ using (\ref{eqn:alphabeta}).
We can also calculate $\hat{D}$ using $\beta$ which is
obtained from the ratio
of the Fourier coefficients ${\cal A}(X,\omega_0)$ and ${\cal A}(X,-\omega_0)$
given by (\ref{eqn:fourierminus},\ref{eqn:fourierplus}):
\begin{equation}
\log \left(\frac{{\cal A}(X,\omega_0)}{{\cal A}(X,-\omega_0)} \right) = 
\log \left(\frac{(-\hat{R}(\hat{Q}_0)-\beta \hat{R}^\prime(\hat{Q}_0))-i \alpha \hat{R}^\prime(\hat{Q}_0) }
                {( \hat{R}(\hat{Q}_0)-\beta \hat{R}^\prime(\hat{Q}_0))+i \alpha \hat{R}^\prime(\hat{Q}_0) } \right) + i 2 \beta X.
\end{equation}
A plot of the imaginary part of this quantity versus $X$ will be linear with
slope $2\beta$ if the effective phase is behaving diffusively.
Figure \ref{fig:beta} shows this plot for the same simulations as presented in
Figure \ref{fig:alpha} and exhibits 
the expected linear behavior near $X_1$ and $X_2$.

The diffusion equation (\ref{eqn:effectiveD}) yields $\alpha=\beta$ 
(cf. (\ref{eqn:alphabeta})) so the slopes indicated on Figure \ref{fig:beta}
should be twice those indicated on Figure \ref{fig:alpha}.
Although the small discrepancies could be due to higher order terms in 
(\ref{eqn:effectiveD}) they are within the margin of error in the measurement
which is mainly due to the short-scale chaotic dynamics.

We conclude from the results in Figures \ref{fig:alpha} and \ref{fig:beta} 
that the dynamics of the effective phase are indeed diffusive
and show in Figure \ref{fig:DvsQ} the wavenumber dependence of 
the effective phase diffusion 
coefficient $\hat{D}(\hat{q})$ along with the analytical phase 
diffusion coefficient for the stationary solution \cite{Ri90a}.
The solid curve shows the analytical phase diffusion coefficient $D$ for the 
stationary solution.
The triangles give the effective phase
diffusion coefficient $\hat{D}$ for the chaotic solutions as measured by using 
the decay rate $\alpha$, while the circles show $\hat{D}$ as measured using 
$\beta$.
Despite the scatter in the data it is clear that $\hat{D}(\hat{q})$ decreases 
with  decreasing $\hat{q}$ and presumably goes to zero at a wavenumber for 
which the diffusion coefficient of the nonchaotic state is still negative as 
shown by the dashed line.
Thus the phase of the system is diffusively unstable over a finite range of 
wavenumbers.
This, along with the conservation of the total phase 
(here the conservation of the total {\it effective} phase),
is known to
lead to stable domains of differing wave numbers, and explains the localization
of the chaotic region in Figure \ref{fig:localized}.

For the wavenumbers at which we were able to measure the effective diffusion
coefficient, the effective phase of the solution remains almost stationary
in the absence of any localized forcing.
For larger wavenumbers, however, the effective phase drifts significantly
on the time scale over which we attempted to make our diffusion coefficient
measurements.
Because of this we were unable to measure the effective diffusion coefficient
for larger wavenumbers than those included in Figure \ref{fig:DvsQ}.

\begin{figure}[p] 
\begin{picture}(420,600)(0,0)
\put(-60,-100) {\includegraphics{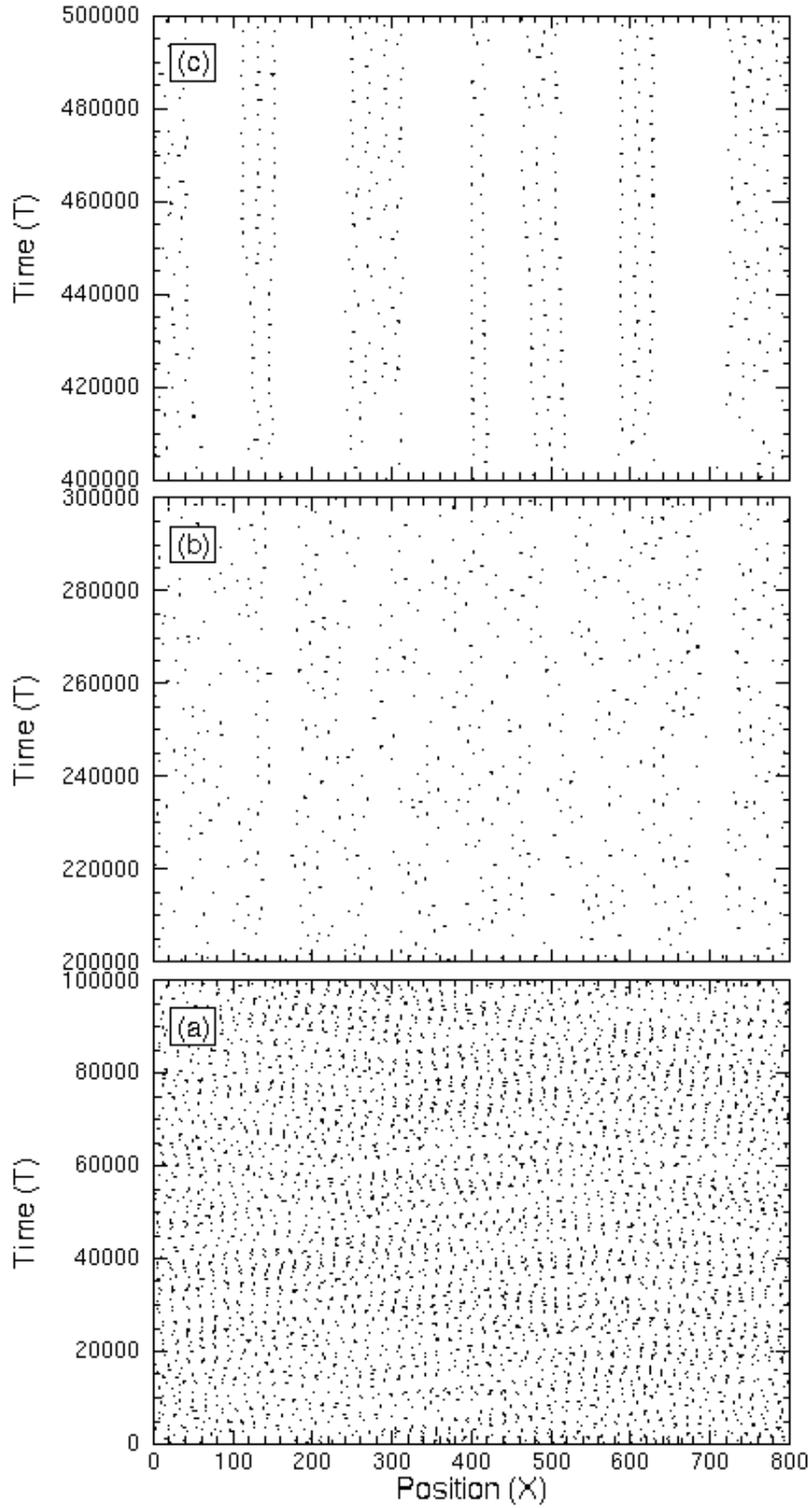}}
\end{picture}
\caption{A set of solutions obtained by incrementally increasing the forcing
amplitude $b$.
The extended spatio-temporal chaos breaks up into localized regions
before the Eckhaus stability limit is reached ($L=800$, $Q=51(2\pi)/L=.4006$).
(a) $b=0.62$, (b) $b=0.64$, (c) $b=0.66$.  
\protect{\label{fig:breakup}}
}
\end{figure}

Figure \ref{fig:DvsQ} suggests that the chaotic activity should not disappear 
homogeneously if the average wavenumber is decreased towards the stable regime.
Instead, the homogeneously chaotic state should 
become unstable to long-wave modulations and eventually 
split up into chaotic and
stationary domains in which the local wavenumber is in the respective stable
regimes ($D>0$ and $\hat{D}>0$).
A similar breaking up of the homogeneous state is expected as the forcing
amplitude $b$ is increased.
Numerical simulations confirm this as shown in Figure \ref{fig:breakup}.

\section{The Two Dimensional Problem} \label{sec:twoD}

In this section we consider a two-dimensional version of 
(\ref{oneDA},\ref{oneDB})
\begin{eqnarray}
\partial_TA+s\partial_XA &=& d\partial_X^2A+d_2\partial_Y^2A+aA+bB
 +cA(|A|^2+|B|^2)+gA|B|^2  
\label{twoD1} \\
\partial_TB-s\partial_XB &=& d^*\partial_X^2B+d_2^*\partial_Y^2B+a^*B+bA
 +c^*B(|A|^2+|B|^2)+g^*B|A|^2 
\label{twoD2}
\end{eqnarray}
These equations describe parametrically excited waves in a two-dimensional 
{\it anisotropic} system.
The usual Faraday experiment is {\it not} described by these equations since 
it is isotropic.
An experimental system that {\it is} described by (\ref{twoD1},\ref{twoD2})
is electroconvection in nematic liquid crystals \cite{ReRa88}, in which the 
anisotropy is due to a preferred direction of the rod-like molecules.
Parametrically excited waves have been observed in both the normal roll regime 
\cite{ReRa88}, in which the waves are perpendicular to the preferred direction
of the molecules, 
and in the oblique roll regime \cite{ToRe90}, in which the waves are at an
oblique angle to the preferred direction.
The normal rolls are described by (\ref{twoD1},\ref{twoD2}) while the oblique
rolls are described by {\it four} coupled amplitude equations \cite{SiRi92}.

The linear stability analysis of the standing wave solution of 
(\ref{twoD1},\ref{twoD2}) must consider
perturbations transverse to the waves as well as along them.
This allows additional instabilities as compared to the one-dimensional case.
For the parameters chosen for the one-dimensional simulations above, along
with the dispersion coefficient in the y-direction $d_2=d$, a stable wavenumber 
band exists for $0.05<b<0.85$.

As in the one-dimensional case (\ref{twoD1},\ref{twoD2}) can be reduced to a 
single Ginzburg-Landau equation with real coefficients near onset.
Therefore the solutions for small $b$ are similar to
those of a single real Ginzburg-Landau equation and the solution with
an initial condition
slightly outside the Eckhaus stability limit undergoes a transition which
reduces the wavenumber.
In two dimensions
the transition from an Eckhaus unstable wave number to a stable
wave number involves the creation of defect pairs which are created together 
and then move apart as shown in Figure \ref{fig:zerocontours}.
There, part of the spatial domain 
of the problem at eight sequential times is represented by the eight squares.
Within each square the solid curves represent locations where the real part of 
the solution is zero, and the dashed curves represent locations where the 
imaginary part of the solution is zero.
Locations where a solid curve crosses a dashed curve are the ``defects'' where 
the amplitude of the solution is zero.
These defects are indicated by circles in Figure \ref{fig:zerocontours}.
The defects, which form as a pair (at some $500<t<600$ in 
Figure \ref{fig:zerocontours}), move apart
leaving one less wave in the domain after they have moved far apart ($t=2500$).
This is analogous to a phase slip in one dimension.

For larger values of $b$ the solutions to (\ref{twoD1},\ref{twoD2}) 
also involve the creation of defect pairs.
However, the defects often do not move far apart but stop after separating 
some finite distance, turn around and annihilate each other.
This can be visualized by considering only the left column of 
Figure \ref{fig:zerocontours}.
The solution begins with the downward sequence of squares from the top
as sequential in time (as in the standard Eckhaus transition).
However, on reaching the bottom of the column ($t=650$) the time sequence
reverses and leads back to the top of the column where the wavenumber resumes
its original value. 
Such a {\it bound defect pair} is analogous to a {\it double} phase slip
in one dimension.
In our two-dimensional simulations we have not found the single 
bound defect pair or simple 
periodic solutions that are suggested by this visualization.
But we have seen solutions where many of the defects which are created
together as pairs annihilate each other.
Figure \ref{fig:defects} is a space-time diagram of such a solution for $b=1.2$.
The spatial aspect of Figure \ref{fig:defects} is a projection; 
the $y$ coordinate is not represented.
Each circle represents the $x$-coordinate of the position 
of a defect and the time it was at this location.
Small and large circles denote defects of opposite polarity (as they do in
Figure \ref{fig:zerocontours}).
Focusing on the ``bubble'' formed by the pair of defects created at 
$x \approx 29$ and $t \approx 2680$ we see that the defects in this pair indeed
move apart for a short time before they reverse their directions, 
ultimately annihilating each other at $x \approx 30$ and $t \approx 2710$.
There are many such bubbles on the diagram as well as
more complex events such as bubbles involving two or three defect pairs.
While for this value of $b$ most defects are ``bound'' in such a bubble,
there are also a few isolated ``unbound'' defects as well.
In these preliminary calculations we have not investigated the statistics of 
bound and unbound defects and the dependence of the statistics on the forcing
amplitude $b$.

Figures \ref{fig:zerocontours} and \ref{fig:defects} are based on $A(X,Y,T)$.
Similar to the phase slips in the one-dimensional problem, 
in our two-dimensional simulations
a defect in $A$ is always accompanied by a defect in $B$ at a nearby location.

\section{Conclusions and Discussion} \label{sec:conclusions}

\begin{figure}[p] 
\begin{picture}(420,270)(0,0)
\put(-50,-50) {\includegraphics{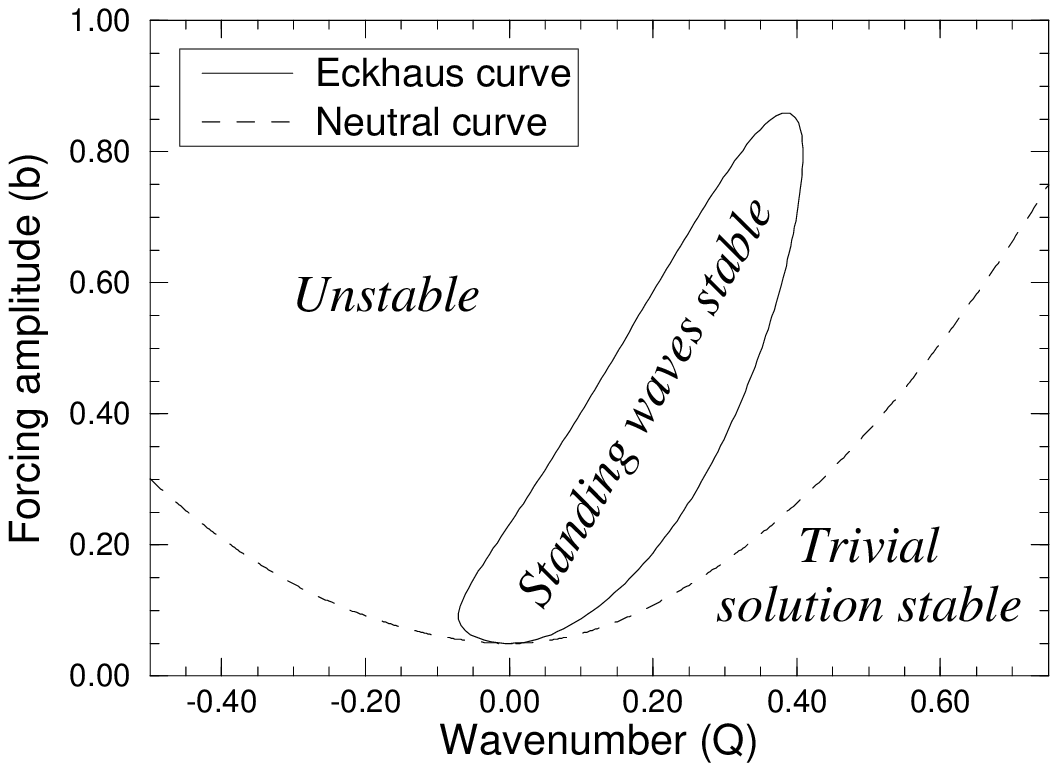}}
\end{picture}
\caption{Linear stability diagram for  
$a=-0.05$, $c=-1+4i$, $d=1+0.5i$, $s=0.2$, $g=-1-12i$.
\protect{\label{fig:linstab}}
}

\begin{picture}(420,310)(0,0)
\put(-70,-250) {\includegraphics{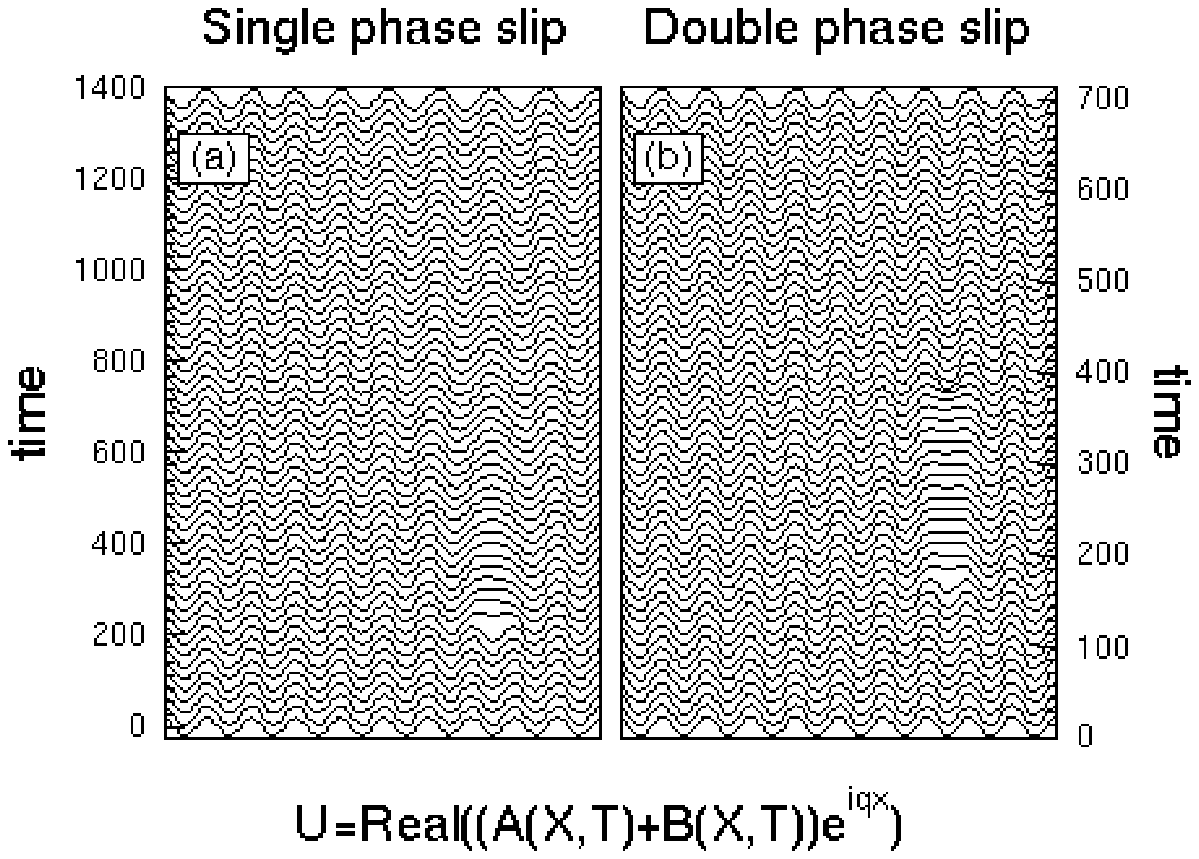}}
\end{picture}
\caption{Space-time diagram illustrating (a) single phase slip observed
for small values of the forcing amplitude (e.g. $b=0.1$), and (b) double
phase slip observed for larger values of the forcing amplitude (e.g. $b=0.6$).
\protect{\label{fig:phaseslip}}
}
\end{figure}

\begin{figure}[p] 
\begin{picture}(420,580)(0,0)
\put(-10,-40) {\includegraphics{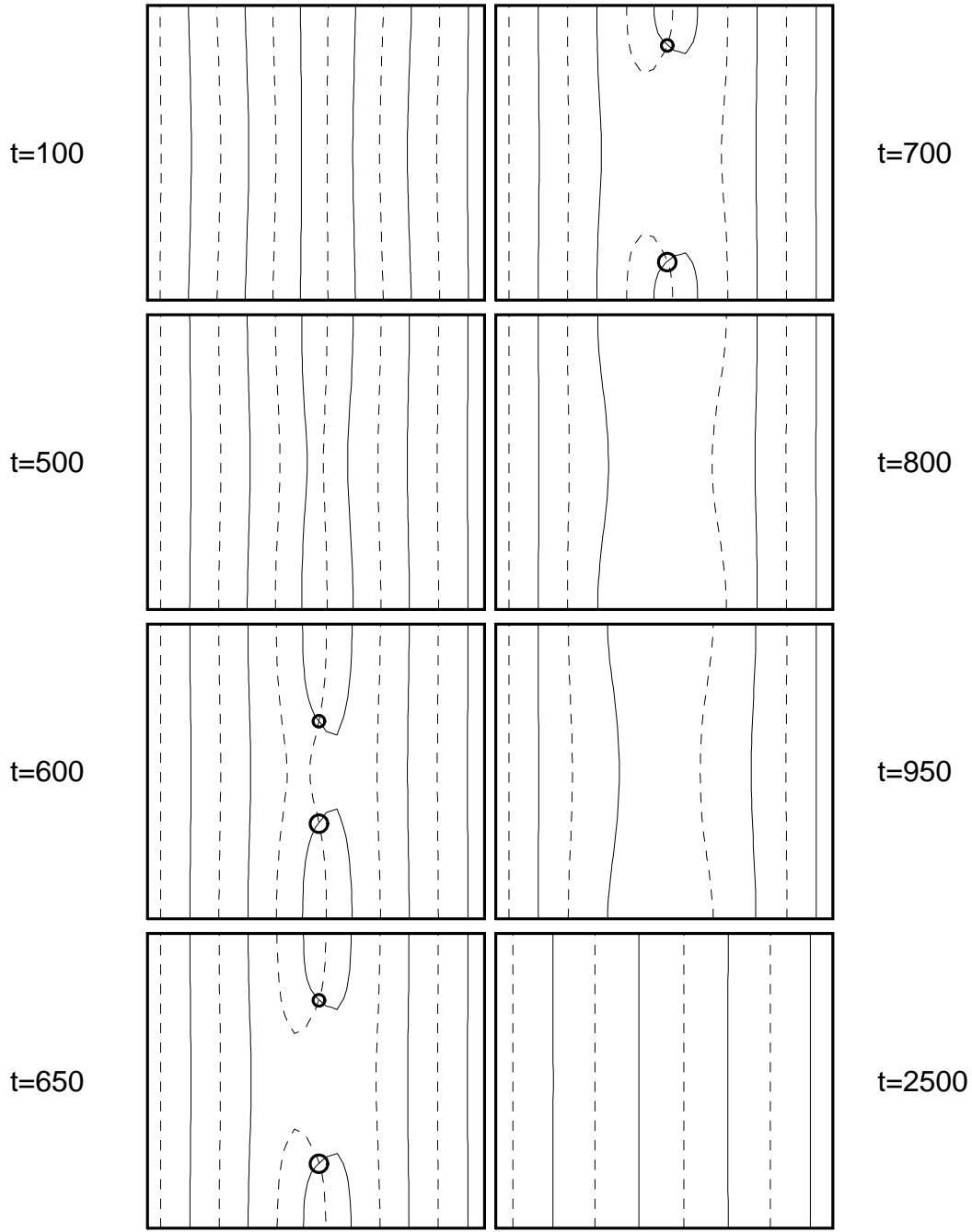}}
\end{picture}
\caption{
Standard wavenumber reducing transition in two dimensions.
Solid curves represent locations where the real part of the solution is zero. 
Dashed curves represent locations where the imaginary part of the solution is zero. 
A {\it bound defect pair} can be visualized by considering the left column
only, progressing from the top to the bottom then returning to the top.
\protect{\label{fig:zerocontours}}
}
\end{figure}

We have presented numerical solutions of a pair of coupled
Ginzburg-Landau equations that are amplitude equations for parametrically
excited waves ((\ref{oneDA},\ref{oneDB}),(\ref{twoD1},\ref{twoD2})).
In particular, we have studied the ensuing dynamics when solutions of these
equations corresponding to standing waves which are phase locked to the 
external forcing are perturbed.
Parameters in the equations were chosen such that there is a 
forcing amplitude above which all wave numbers are linearly unstable 
(Figure \ref{fig:linstab}).
The usual transition from an Eckhaus-unstable state to a stable state 
(Figures \ref{fig:phaseslip}a, \ref{fig:zerocontours}) is found near onset but 
not for larger forcing amplitudes.

\begin{figure}[p] 
\begin{picture}(420,270)(0,0)
\put(-50,-50) {\includegraphics{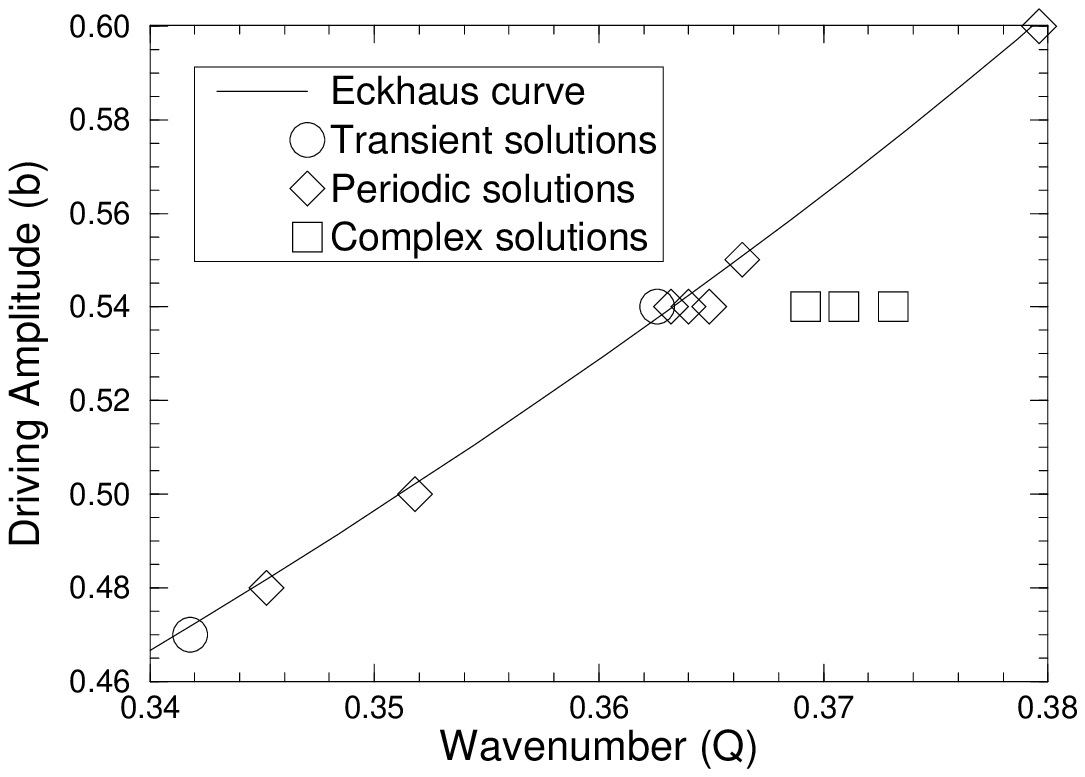}}
\end{picture}
\caption{Example values of the forcing amplitude $b$ and wavenumber $Q$ where 
the three solution types occur.
\protect{\label{fig:solutions}}
}

\begin{picture}(420,300)(0,0)
\put(-50,-50) {\includegraphics{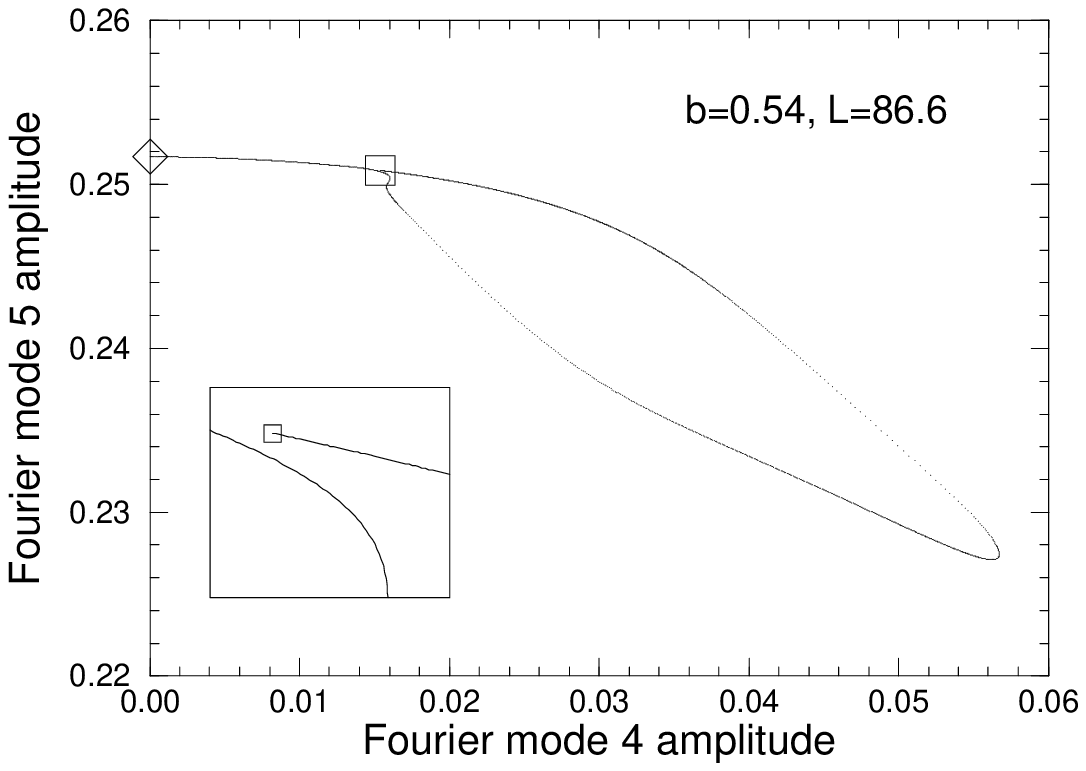}}
\end{picture}
\caption{Phase plane representation of a transient solution involving 
one double phase slip.
The square indicates the initial condition which is the perturbed
5-wave state.
The loop is traversed once, with the final state of the solution 
being the (stable) fixed point indicated by a diamond.
($b= 0.54$, $Q=5(2\pi)/L= 0.3628$).
\protect{\label{fig:phaseplane1DPS}}
}
\end{figure}

\begin{figure}[p] 
\begin{picture}(420,270)(0,0)
\put(-50,-50) {\includegraphics{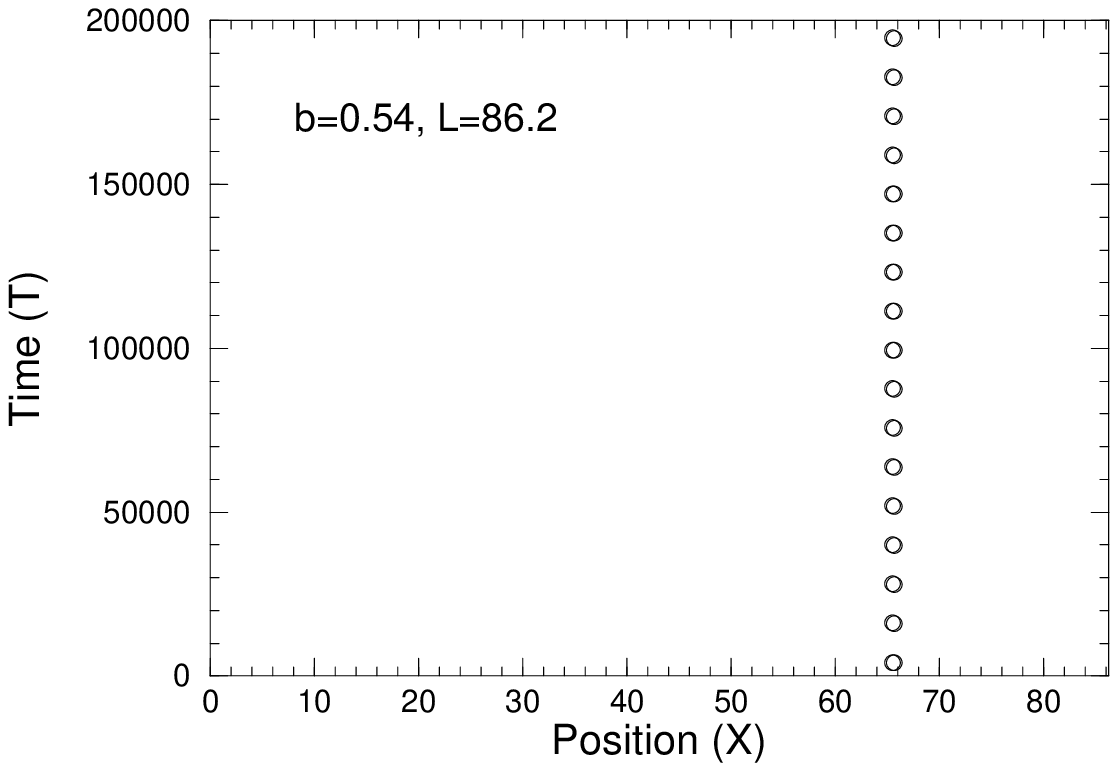}}
\end{picture}
\caption{Each circle represents the location (in space and time) of a double 
phase slip for a simple periodic solution
($b=0.54$, $Q=5(2\pi)/L=0.3645$).
\protect{\label{fig:simpleperiodic}}
}

\begin{picture}(420,300)(0,0)
\put(-50,-50) {\includegraphics{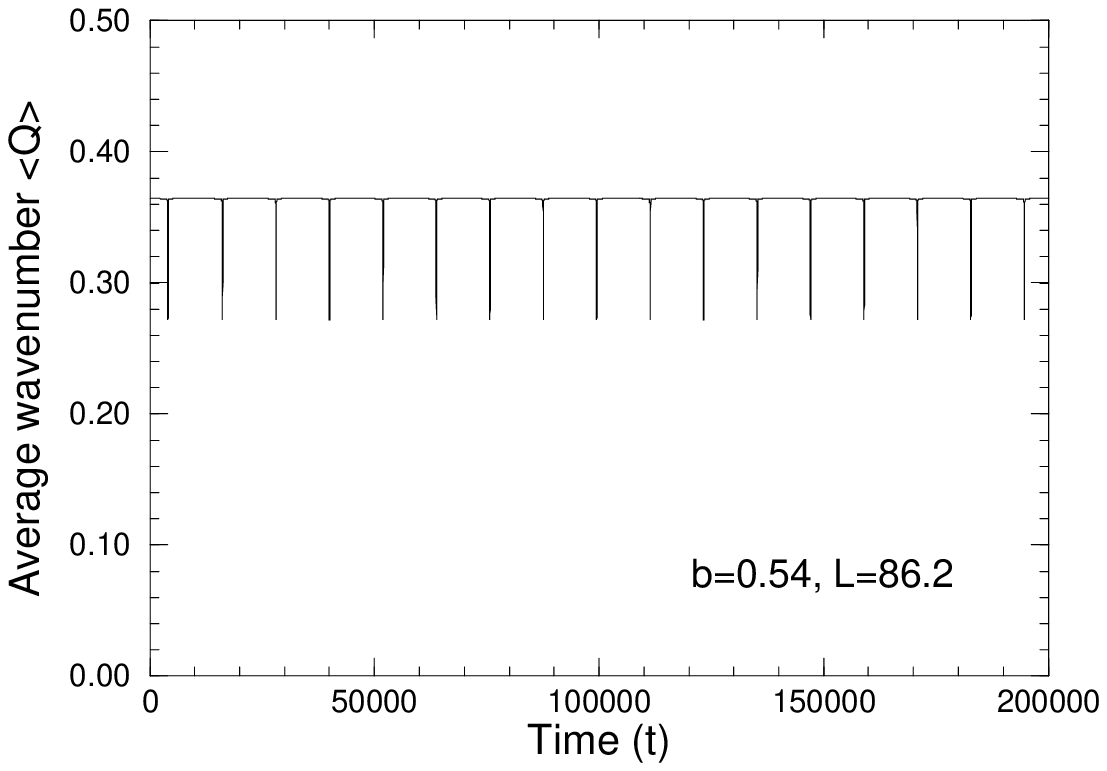}}
\end{picture}
\caption{Spatial average of the local wavenumber plotted as a function of time
for a simple periodic solution
($b=0.54$, $Q=5(2\pi)/L=0.3645$).
\protect{\label{fig:qvstime}}
}
\end{figure}

\begin{figure}[p] 
\begin{picture}(420,270)(0,0)
\put(-82,-270) {\includegraphics{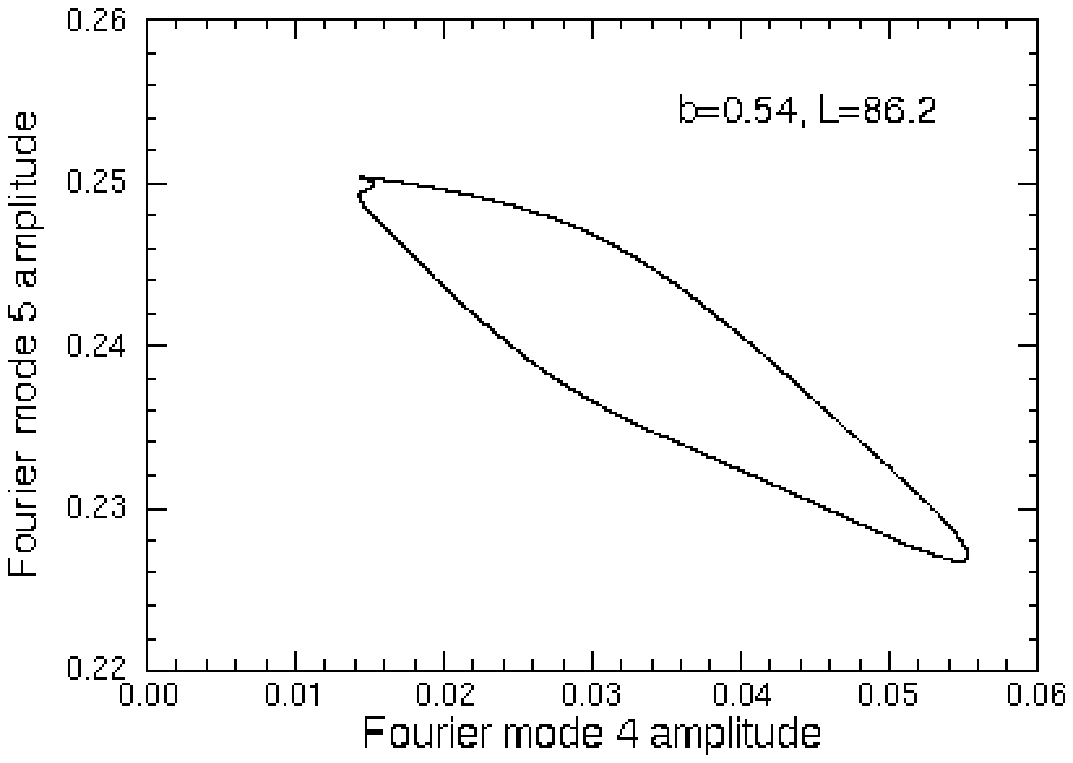}}
\end{picture}
\caption{Phase plane representation of a simple periodic solution
($b=0.54$, $Q=5(2\pi)/L=0.3645$).
\protect{\label{fig:phaseplaneSPS}}
}

\begin{picture}(420,300)(0,0)
\put(-50,-50) {\includegraphics{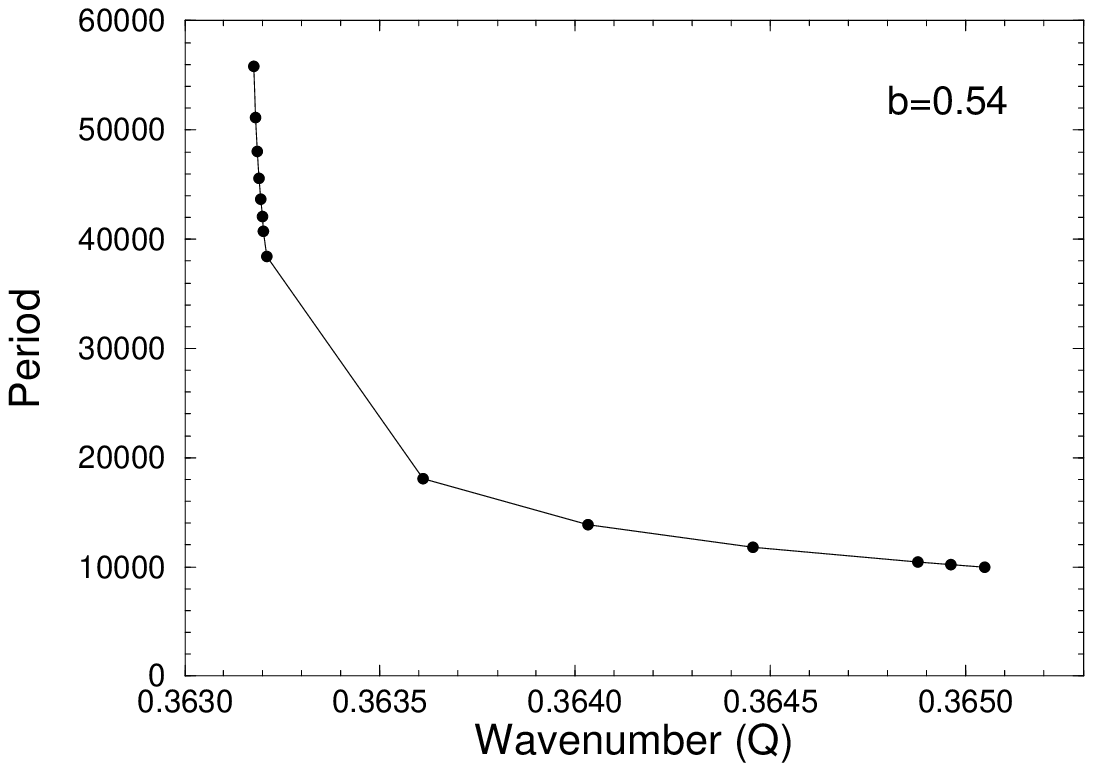}}
\end{picture}
\caption{Divergence of the period of the simple periodic solutions as
the wavenumber $Q$ is decreased
($b=0.54$).
\protect{\label{fig:period}}
}
\end{figure}

\begin{figure}[p] 
\begin{picture}(420,270)(0,0)
\put(-50,-50) {\includegraphics{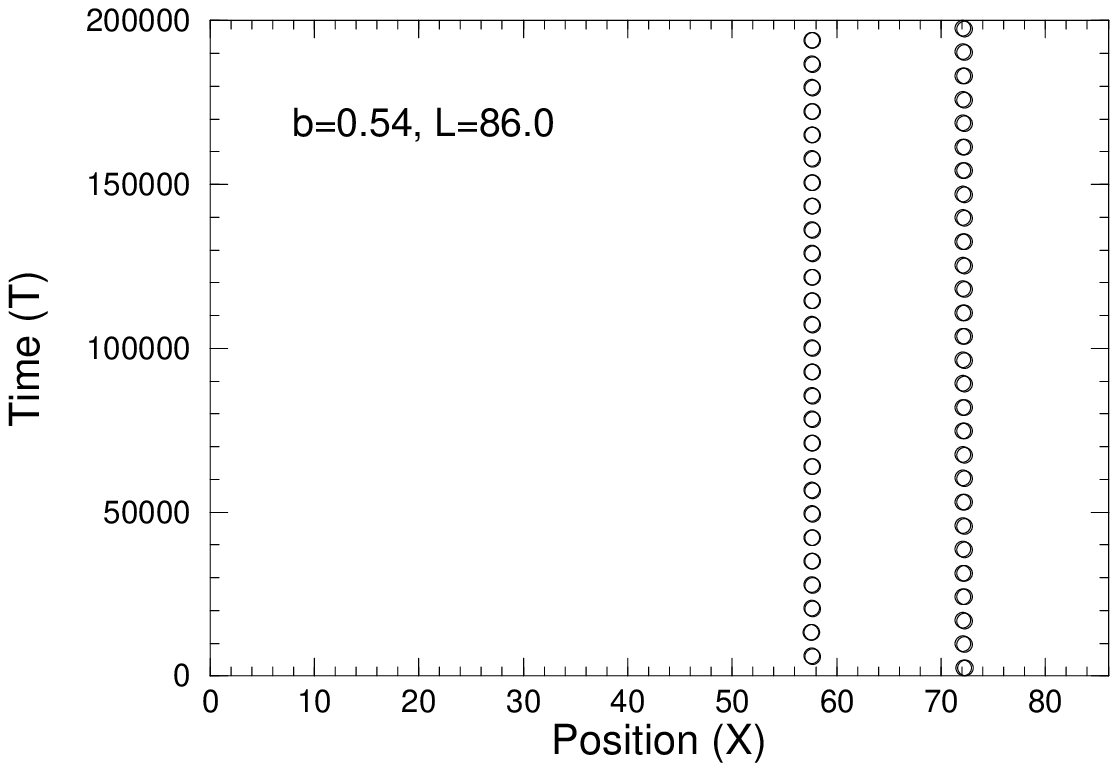}}
\end{picture}
\caption{Location of double phase slips for a periodic solution occurring after 
a period doubling bifurcation
($b=0.54$, $Q=5(2\pi)/L=0.3653$).
\protect{\label{fig:perioddoubled}}
}

\begin{picture}(420,300)(0,0)
\put(-50,-50) {\includegraphics{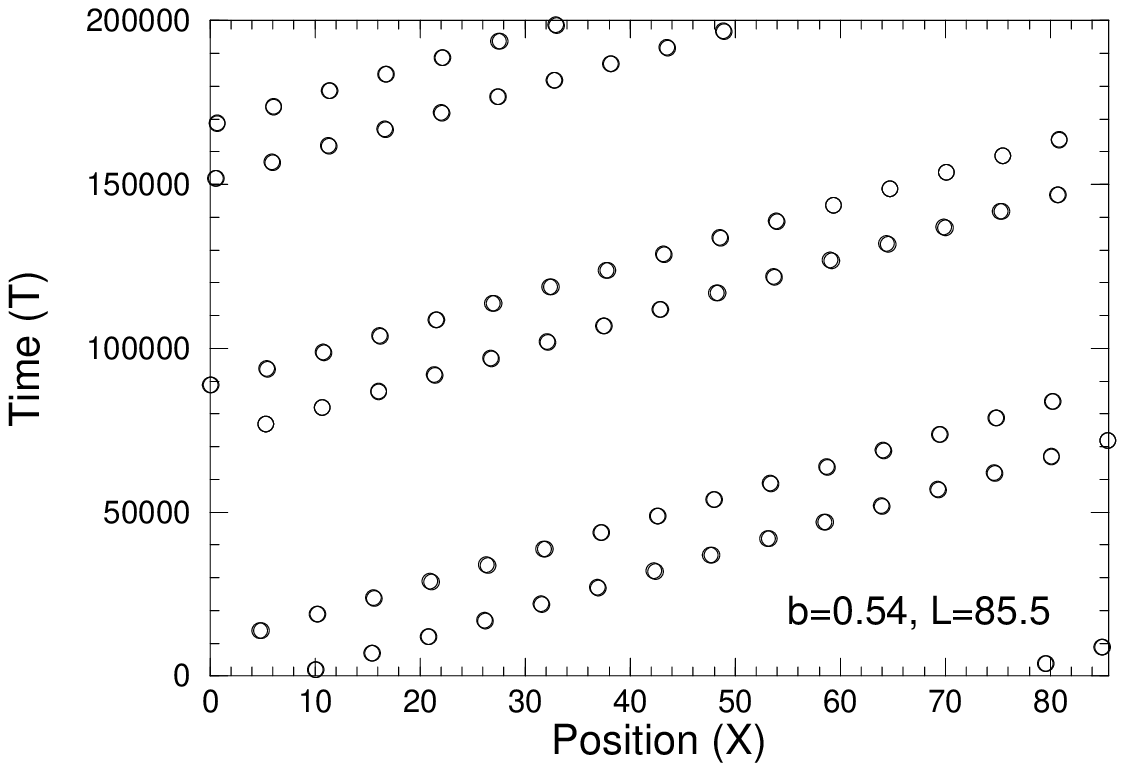}}
\end{picture}
\caption{Location of double phase slips for a solution where a wave of double 
phase slips propagates through the system
($b=0.54$, $Q=5(2\pi)/L=0.3674$).
\protect{\label{fig:travelingwave}}
}
\end{figure}

In one dimension {\it double phase slips} (Figure \ref{fig:phaseslip}b)
occur with these larger forcing amplitudes.
These double phase slips allow periodic dynamics with 
an average wavenumber that is Eckhaus unstable even for parameters for which 
there is a range of stable wavenumbers 
(Figures \ref{fig:solutions}, \ref{fig:simpleperiodic}-\ref{fig:travelingwave}).
Solutions with an average wave number that is stable can also exhibit 
transient and periodic solutions that include double phase slips.  
In this regime the system can be excitable (Figure \ref{fig:phaseplane1DPS}).

\begin{figure}[p] 
\begin{picture}(420,450)(0,0)
\put(-50,-50) {\includegraphics{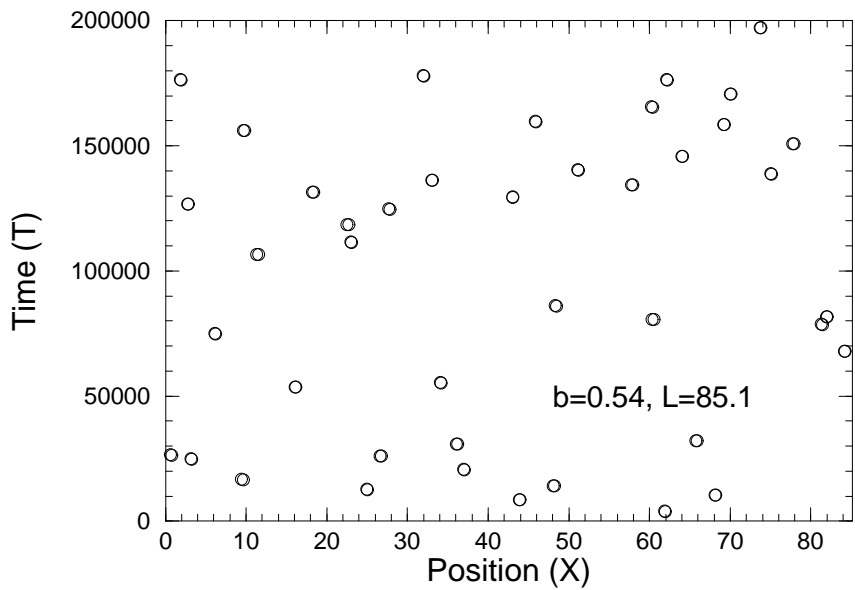}}
\end{picture}
\caption{A complex solution where the double phase slips occur irregularly in 
space and time
($b=0.54$, $Q=5(2\pi)/L=.3692$). 
\protect{\label{fig:complex}}
}
\end{figure}

\begin{figure}[p] 
\begin{picture}(420,600)(0,0)
\put(-60,-100) {\includegraphics{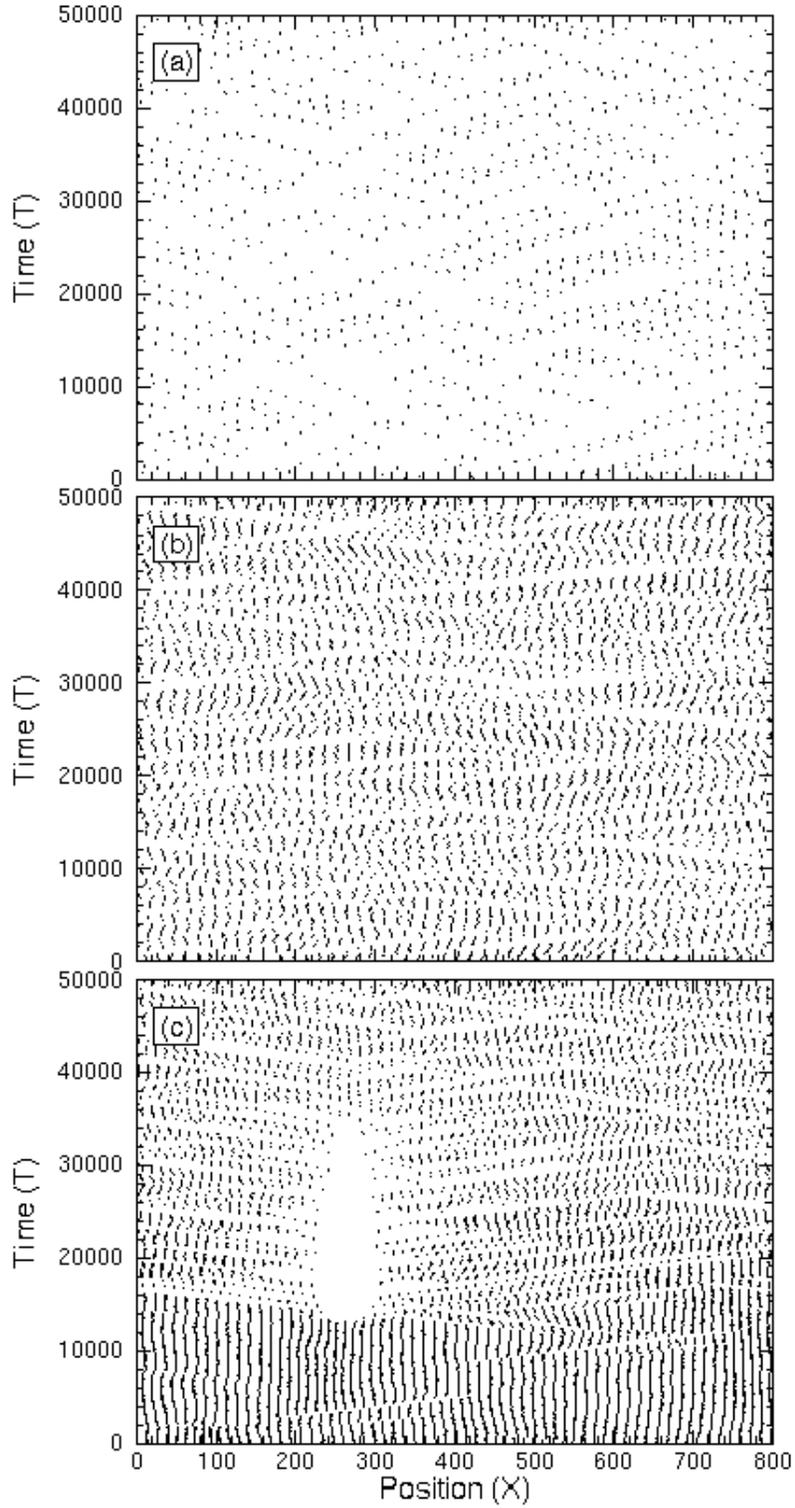}}
\end{picture}
\caption{Extended spatio-temporal chaos in a large system ($b=0.6$, $L=800$):
(a) $Q=50(2\pi)/L=0.3927$, 
(b) $Q=51(2\pi)/L=0.4006$, 
(c) Initial wavenumber $Q_0=52(2\pi)/L=0.4084$, after a single phase slip at
$t\approx13000$ the wavenumber is $Q=0.4006$.
\protect{\label{fig:extended}}
}
\end{figure}

\begin{figure}[p] 
\begin{picture}(420,270)(0,0)
\put(-100,-250) {\includegraphics{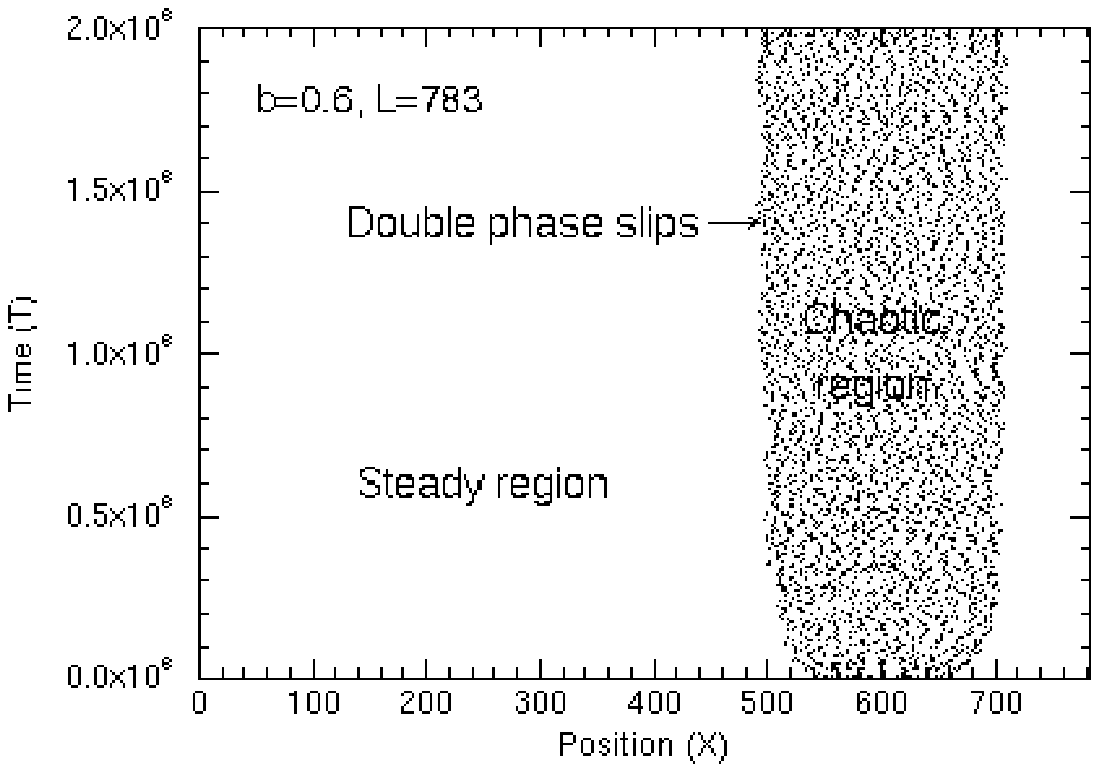}}
\end{picture}
\caption{Localized spatio-temporal chaos in a large domain.
Each dot represents a double phase slip
($b=0.6$, $L=783$, $Q=47(2\pi)/L=0.377$).
\protect{\label{fig:localized}}
}

\begin{picture}(420,300)(0,0)
\put(-50,-50) {\includegraphics{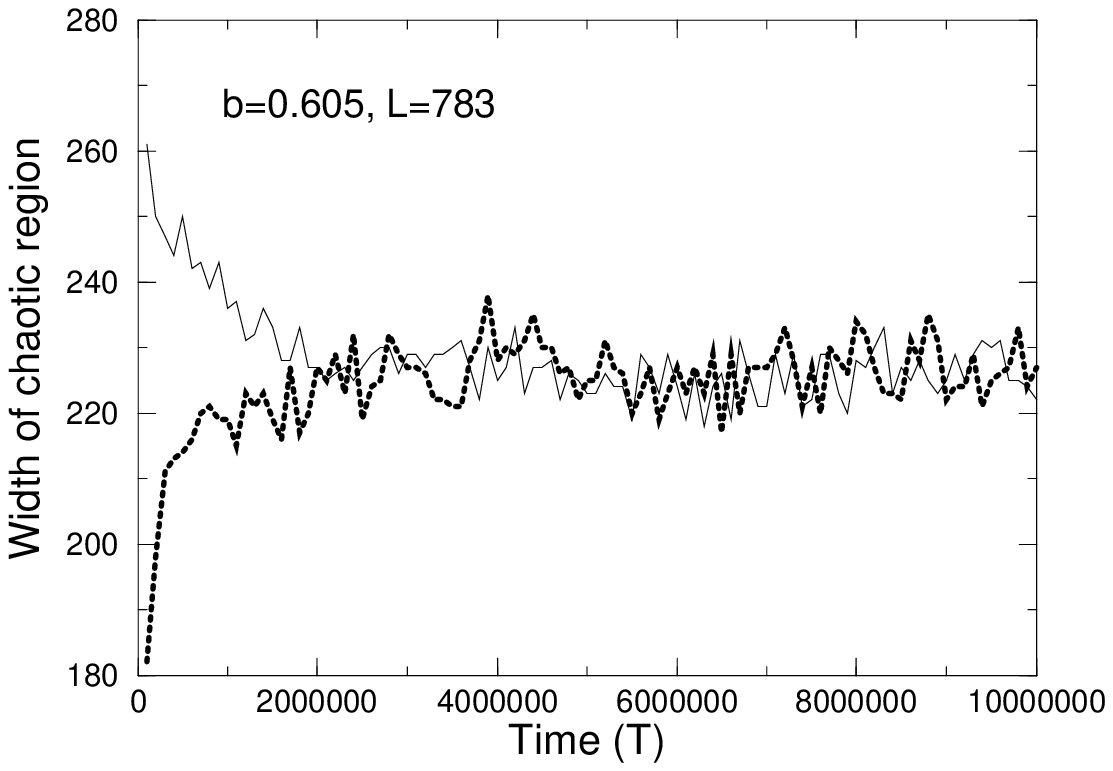}}
\end{picture}
\caption{Temporal evolution of the width of the localized region exhibiting 
spatio-temporal chaos for two different initial conditions
($b=0.605$, $Q=47(2\pi)/L=0.377$).
\protect{\label{fig:width}}
}
\end{figure}

\begin{figure}[p] 
\begin{picture}(420,270)(0,0)
\put(-50,-50) {\includegraphics{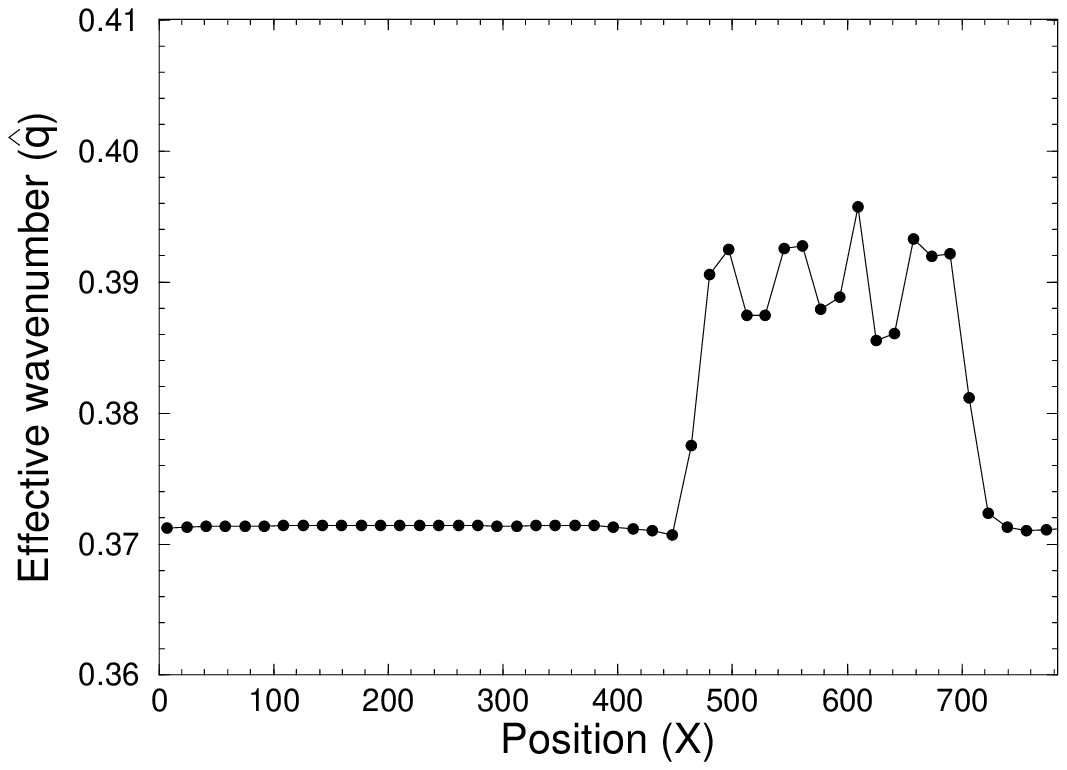}}
\end{picture}
\caption{Local wavenumber of the time average of $A(X,T)$ for a localized 
chaotic solution.
The local wavenumber in the chaotic region is larger than in the quiescent 
region.
($b=0.6$, $L=783$, $Q=47(2\pi)/L=0.377$).
\protect{\label{fig:wavenumber}}
}
 
\begin{picture}(420,300)(0,0)
\put(-50,-50) {\includegraphics{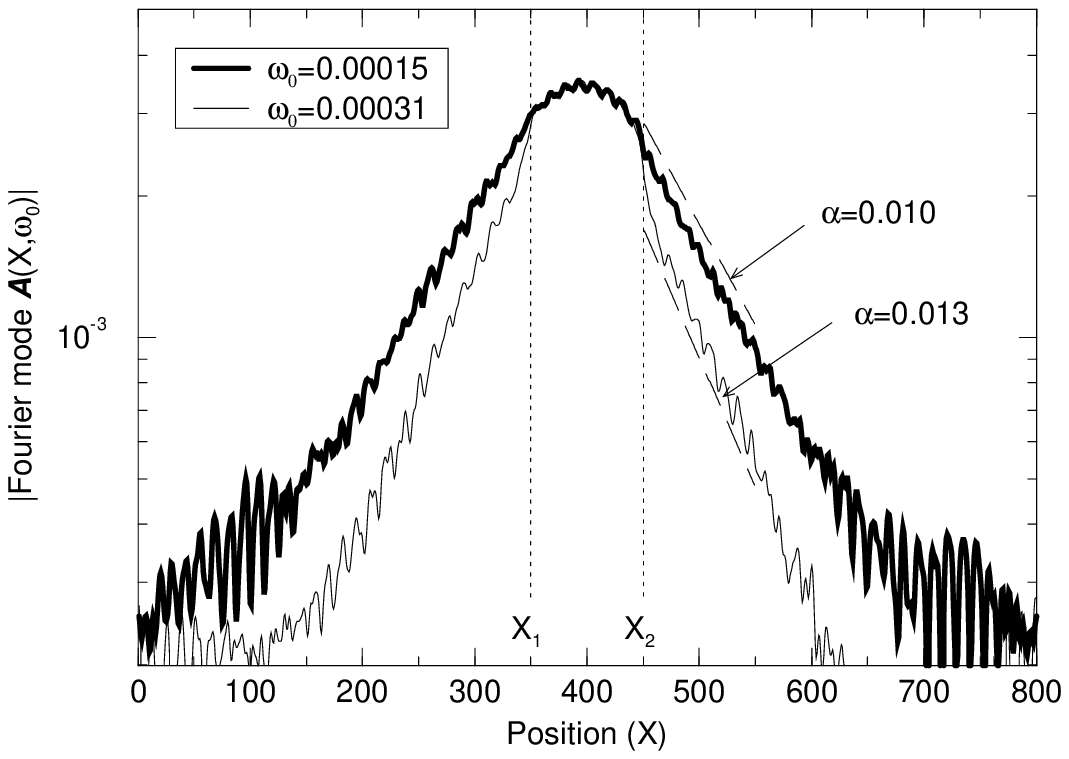}}
\end{picture}
\caption{Decay of the magnitude of the coefficient of the time Fourier transform
of $A$ corresponding to the frequency $\omega_0$ in the localized advection 
term $v$ in equations (\protect{\ref{eqn:advectA}},\protect{\ref{eqn:advectB}}), 
for two different values of $\omega_0$ 
($b=0.61$, $L=800$, $Q=51(2\pi)/L= 0.4006$, $v_0=1.630\omega_0$).
\protect{\label{fig:alpha}}
}
\end{figure}

\begin{figure}[p] 
\begin{picture}(420,270)(0,0)
\put(-50,-50) {\includegraphics{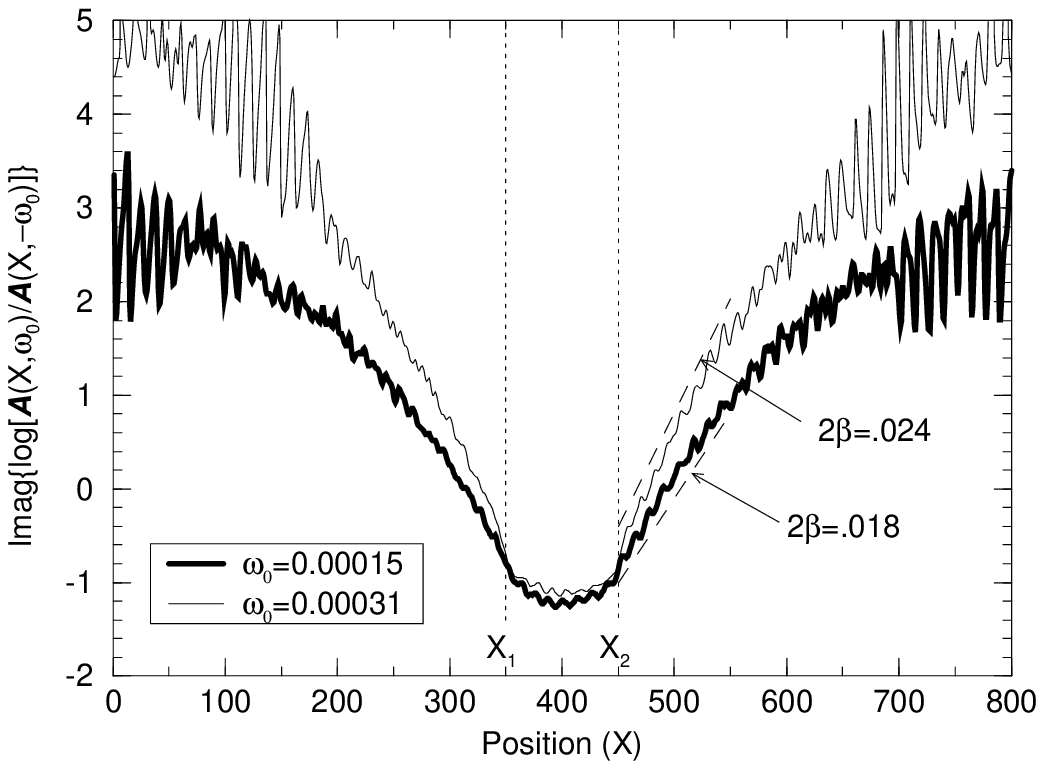}}
\end{picture}
\caption{Imaginary part of the ratio of the coefficients of the time Fourier 
transform of A corresponding to the frequencies $\pm\omega_0$ in the localized 
advection term $v$ in equations 
(\protect{\ref{eqn:advectA}},\protect{\ref{eqn:advectB}}),
for two different values of $\omega_0$ 
($b=0.61$, $L=800$, $Q=51(2\pi)/L= 0.4006$, $v_0=1.630\omega_0$).
\protect{\label{fig:beta}}
}

\begin{picture}(420,300)(0,0)
\put(-50,-50) {\includegraphics{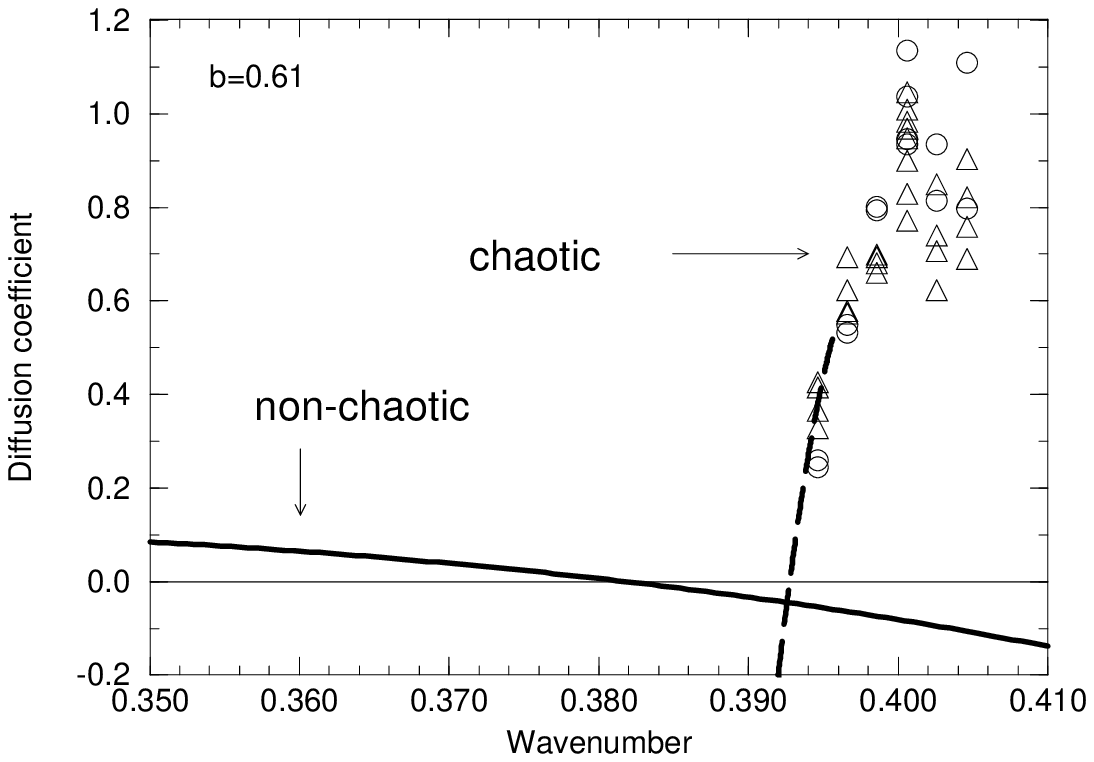}}
\end{picture}
\caption{Analytical phase diffusion coefficient $D$ for the stationary solution
(solid line) and effective diffusion coefficient $\hat{D}$ for the chaotic 
solution;  Triangles and circles represent numerical results as obtained from
$\alpha$ and $\beta$ (cf. Figures \protect{\ref{fig:alpha}} and 
\protect{\ref{fig:beta}}), the dashed line 
is the presumed extension of these results ($b=0.61$).
\protect{\label{fig:DvsQ}}
}
\end{figure}

Also found in one dimension are solutions where the dynamics are chaotic and 
characterized by the irregular occurrence of double phase slips in both space 
and time (Figures \ref{fig:complex}, \ref{fig:extended}). 
Particularly striking are solutions in which the chaos is confined to only a 
part of the homogeneous system 
(Figures \ref{fig:localized}-\ref{fig:wavenumber}).
An effective phase diffusion equation utilizing the long-term phase 
conservation of the solution explains the localization of this 
new form of amplitude chaos (Figures \ref{fig:alpha}-\ref{fig:DvsQ}).

\begin{figure}[p] 
\begin{picture}(420,500)(0,0)
\put(-50,-250) {\includegraphics{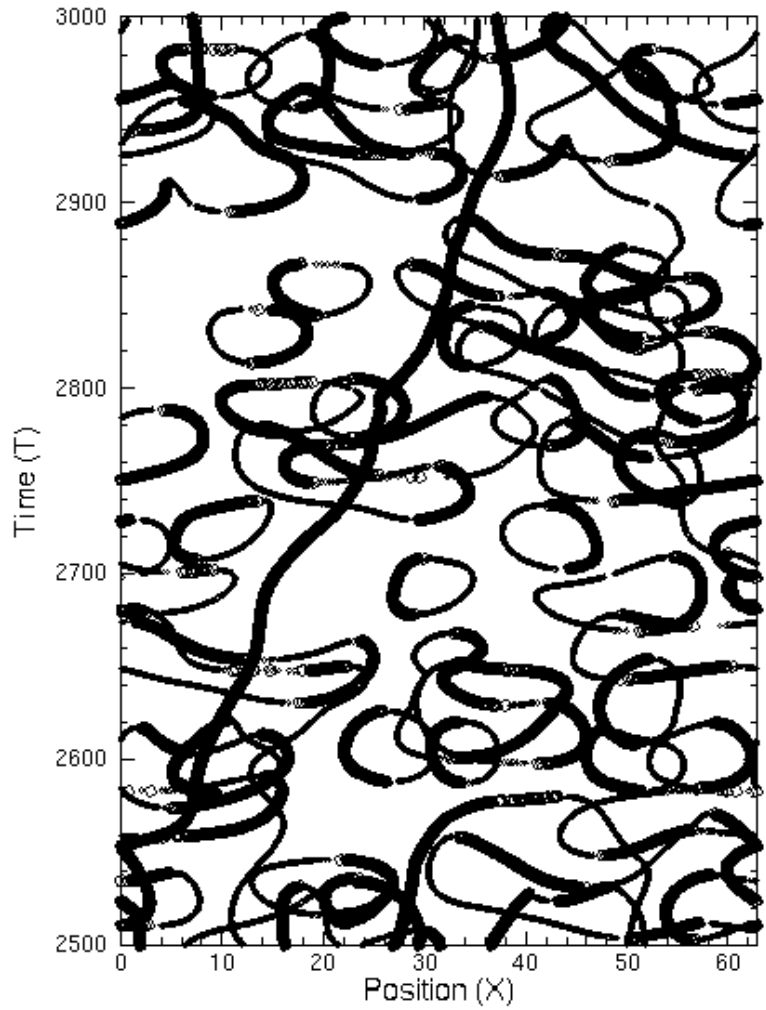}}
\end{picture}
\caption{
Space-time diagram showing the projection of the location of defects on the
$x$-axis for a simulation of the two-dimensional system.
The {\it bubbles}, for example the one starting at $x \approx 29$ and 
$t \approx 2680$, represent bound defect pairs ($b=1.2$, $L=62.832$).
\protect{\label{fig:defects}}
}
\end{figure}

A double phase slip in one dimension corresponds to a {\it bound defect pair} in
two dimensions.
A bound defect pair consists of two dislocations which are formed together, 
move apart, then move back together and annihilate each other.
These bound defect pairs are prominent in our two-dimensional simulations 
(Figure \ref{fig:defects}).

So far we have not identified a particular physical system which exhibits the
behavior described here.
It is natural to suspect that the closing of the Eckhaus curve 
(cf. Figure \ref{fig:linstab}) is 
related to the occurrence of double phase slips.
If that should be the case, parametric driving of waves which are Benjamin-Feir 
unstable in the absence of any forcing should be a good candidate for this 
behavior.
Within the framework of the coupled Ginzburg-Landau equations (1,2) 
these waves become unstable near the band 
center when forced sufficiently strongly \cite{Ri90}.
Depending on the sign of certain nonlinear coefficients this indicates 
instability at all wavenumbers and a closing of the Eckhaus curve.
Further work is needed to investigate the origin of the double phase slips and 
parameter values for which the 
dynamics seen in our study can be observed.
Of particular interest is whether double phase slips might occur in a Faraday
experiment.

Independent of experimental realizability, a very interesting theoretical
question arises with regard to the bound defect pairs in two dimensions.
In thermodynamic equilibrium the unbinding transition of defect pairs has been
studied in great detail in the context of two-dimensional melting 
\cite{NeHa79,KoTh73}.
In the vicinity of that phase transition quite unusual scaling properties were 
found.
It has been conjectured that such a transition should also be possible in 
nonequilibrium systems \cite{OcGu83,WaDe82,CrHo93} but no very convincing 
instance has been identified so far.

The use of an effective phase diffusion equation for the large scale behavior 
of the chaotic state where the effective diffusion coefficient is positive but 
the usual phase diffusion coefficient is negative (indicating that the 
non-chaotic solution is unstable) is similar to studies using the 
Kardar-Parisi-Zhang equation \cite{KaPa86}, which is Burger's equation with a 
Gaussian forcing term added, to model chaotic solutions of the 
Kuramoto-Sivashinsky equation \cite{Ya81,Za89,JaHa93}.
Presumably a similar noise term $\zeta(X,T)$ could be added to 
(\ref{eqn:effectiveD}) to 
describe the effect of the shortscale chaotic dynamics on the effective phase,
\begin{equation}
\partial_T \hat{\phi} = \hat{D}_0 \partial^2_X \hat{\phi} + 
                       \hat{D}_1 \partial_X \hat{\phi} \partial^2_X \hat{\phi} +
                       \zeta(X,T) + \ldots.  \label{eqn:noisy}
\end{equation}
The form of the nonlinear term, which results from expanding the diffusion 
coefficient $\hat{D}$ around some wavenumber $\hat{q}_0$, differs from that 
found in the Kardar-Parisi-Zhang equation since the  
parametrically driven standing waves have an additional reflection symmetry
as compared to the traveling 
waves described by the Kuramoto-Sivashinsky equation.
Power counting \cite{Am78} indicates that in the limit of very long scales the
nonlinear terms in (\ref{eqn:noisy}) are not relevant (neither in one
nor two dimensions), in contrast to the nonlinear terms in the 
Kardar-Parisi-Zhang equation.
A particularly interesting feature of  
(\ref{eqn:noisy}) is, however, that even the effective diffusion 
coefficient $\hat{D}_0$ can become negative.
This is not reported for the Kardar-Parisi-Zhang equation as applied to the 
Kuramoto-Sivashinsky equation.
It suggests that noise may be quite important in the regime in which $\hat{D}$ 
is small, i.e. near the onset of domain formation.

We would like to thank J. Eggers, G. Grinstein, and Y. Tu for stimulating 
discussions.
This work was supported by the United States Department of Energy through 
grant DE-FG02-92ER14303.


\bibliography{/home2/hermann/.index/journal}

\pagebreak


\end{document}